\documentclass[a4paper,12pt]{article}
 
\PassOptionsToPackage{original}{pict2e}
\usepackage{amsmath}
\usepackage{amssymb}
\usepackage{amsfonts}
\usepackage{mathrsfs}
\usepackage{bm}
\usepackage{mathtools}
\usepackage{geometry}
\usepackage{dsfont}
\usepackage{lmodern}
\usepackage{calc}
\usepackage[english]{babel}
\usepackage{color}
\usepackage{xcolor}
\usepackage{upgreek}
\usepackage{graphicx,pspicture}
\usepackage{subfig}
\usepackage{psfrag}
\usepackage{afterpage}
\usepackage{float}
\usepackage{array}
\usepackage{tabularx}
\usepackage{arydshln}
\usepackage{dashrule}
\usepackage[margin=10pt,font=small,labelfont=bf,labelsep=period]{caption}
\usepackage{upref} 
\usepackage{cite} 
\usepackage[bottom]{footmisc} 
\usepackage{indentfirst} 
\usepackage{textcomp}
 
\definecolor{rhsFormColor}{rgb}{0.05,0.05,0.35}
\definecolor{Formel}{rgb}{0.05,0.05,0.35}
\definecolor{Formel2}{rgb}{0.40,0.40,0.55}
\definecolor{SideNote}{rgb}{0.35,0.35,0.4}
\definecolor{fSubDeriv}{rgb}{0.1,0.1,0.1}
\definecolor{fSubDerivA}{rgb}{0.35,0.35,0.50}
\definecolor{Text}{rgb}{0,0,0}
\definecolor{can1}{rgb}{0.7,0.1,0.1}
\definecolor{can2}{rgb}{0.45,0.45,0.15}
\definecolor{can3}{rgb}{0.05,0.3,0.05}
\definecolor{can4}{rgb}{0.5,0.1,0.1}
\definecolor{can5}{rgb}{0.1,0.5,0.1}
\definecolor{can6}{rgb}{0.15,0.45,0.45}
\definecolor{can7}{rgb}{0.3,0.05,0.05}
\definecolor{can8}{rgb}{0.35,0.1,0.35}
\definecolor{can9}{rgb}{0.5,0.3,0.4}
\definecolor{SumWeylCol}{rgb}{0.05,0.05,0.35}
\definecolor{SumGeoCol}{rgb}{0.05,0.05,0.35}
\colorlet{TableR1}{Formel}
\colorlet{TableR2}{Formel!75}
\colorlet{TableC1}{SideNote}
\definecolor{TableC2}{rgb}{0.8,0.8,0.8}
\colorlet{TableC3}{SideNote!10}
\definecolor{TableNot}{rgb}{0.8,0.8,0.8}
\definecolor{TableYes}{rgb}{1,1,1}
\definecolor{pageNumberC}{rgb}{0,0,0}

\numberwithin{equation}{section}

\renewcommand{\baselinestretch}{1.2}
\include{globalDefinitions} 

\begin{document}

\addtolength{\topmargin}{-10mm} 
\begin{titlepage}
\enlargethispage{1cm}
\renewcommand{\baselinestretch}{1.1} 
\title{\begin{flushright}
\normalsize{MITP\slash 14-050}
\vspace{1cm}
\end{flushright}
Propagating gravitons vs. `dark matter' in\\
asymptotically safe quantum gravity}
\date{}
\author{Daniel Becker and Martin Reuter\\
{\small Institute of Physics, University of Mainz}\\[-0.2cm]
{\small Staudingerweg 7, D-55099 Mainz, Germany}}
\maketitle\thispagestyle{empty}

\begin{abstract} 
 Within the Asymptotic Safety scenario, we discuss whether Quantum Einstein Gravity (QEG) can give rise to a semi-classical regime of propagating physical gravitons (gravitational waves) governed by an effective theory which complies with the standard rules of local quantum field theory.
According to earlier investigations based on single-metric truncations there is a tension between this requirement and the condition of Asymptotic Safety since the former (latter) requires a positive (negative) anomalous dimension of Newton's constant.
We show that the problem disappears using the bi-metric renormalization group flows that became available recently:
They admit an asymptotically safe UV limit and, at the same time, a genuine semi-classical regime with a positive anomalous dimension.
This brings the gravitons of QEG on a par with arbitrary (standard model, etc.) particles which exist as asymptotic states.
We also argue that metric perturbations on almost Planckian scales might not be propagating, and we propose an interpretation as a form of `dark matter'.
\end{abstract}

\end{titlepage}

\section{Introduction}
One of the indispensable requirements an acceptable fundamental quantum gravity theory must satisfy is the emergence of a classical regime where in particular small perturbations, i.e. gravitational waves, can propagate on an almost flat background spacetime. This regime should be well described by classical General Relativity or, if one pushes its boundary towards the quantum domain a bit further,  by the effective quantum field theory approach pioneered by Donoghue \cite{donoghueEQFT}. 

In this paper we shall consider the scenario where the ultraviolet (UV) completion of quantized gravity is described by an asymptotically safe quantum field theory \cite{wein}.
In a formulation based upon the gravitational average action \cite{mr}, this quantum field theory is defined by a specific renormalization group (RG) trajectory $k\mapsto\EAA_k$ which lies entirely within the UV-critical hypersurface of a non-Gaussian fixed point (NGFP).
Here $\EAA_k\equiv \EAA_k[\flcb_{\mu\nu};\bg_{\alpha\beta}]$ denotes the Effective Average Action, a `running' action functional  which, besides the scale $k$, depends on the (expectation value of the) metric fluctuations, $\flcb_{\mu\nu}$, and the metric of the background spacetime on which they are quantized, $\bg_{\alpha\beta}$.

To recover classical General Relativity in this setting it would be  most natural  if the asymptotically safe RG trajectory of the fundamental theory, emanating from the NGFP in the UV ($k\rightarrow\infty$), contains a segment in the low energy domain ($k\rightarrow 0$) where the full fledged description in terms of the effective average action, valid for all scales and all backgrounds, smoothly goes over into the effective field theory of spin-2 quanta propagating on a rigid background Minkowski spacetime. 
The simplest picture would then be that the approximating low energy theory which is implied by the fundamental asymptotically safe one is `standard' in the sense that it complies with the usual axiomatics of local quantum field theory on Minkowski space which underlies all of particle physics, for instance. 

However, almost all existing RG studies of the Asymptotic Safety scenario, using functional RG methods, indicate that there is a severe tension, if not a clash, between their predictions and the picture of a {\it conventional Minkowski space theory} describing propagating gravitons or gravitational waves at low energies \cite{frank1,frank2,frank+friends,frank-sig,frankfrac,frank-fR,frank-ghost,astrid-ghost,oliver1,oliver2,oliver3,oliver4,oliverfrac,perper,prop,elisa1,vacca,max-pert,creh1,percacci,percadou,percacci-pagani,codello,JE1,JEUM,JEe-omega,QEG+QED,andi1,daniel1,daniel-MG,NJP,livrev}.

In the following we try to describe this tension as precisely as possible. 
It is necessary to distinguish the real question of (non-)existing propagating gravitational waves in the classical regime from  certain objections against Asymptotic Safety in general that were raised occasionally but were based on  misconceptions and are unfounded therefore.
One of these misconceptions is the believe that the anomalous dimensions of quantum fields  must be {\it positive, always.}

In fact, for asymptotically safe Quantum Einstein Gravity (QEG) it is crucial that the anomalous dimension of the metric fluctuations, $\eta_{\text{N}}$, is {\it negative}, at least in the vicinity of the NGFP.
There, by the very construction of the theory's UV completion, it assumes the value $\eta_{\text{N}}^*=-(d-2)$, in $d$ spacetime dimensions.%
\footnote{We assume $d>2$ throughout.} 
And indeed, the RG equations obtained within the special class of non-perturbative approximations that have been considered in the past almost exclusively, the so called `single-metric' truncations of theory space, had always given rise to  a negative anomalous dimension \cite{NJP,livrev}.
Moreover, $\eta_{\text{N}}<0$ was found not only near the NGFP but even {\it everywhere} on the truncated theory space considered. 

In these truncations the ansatz for the Effective Average Action (EAA) always included a term $\propto G_k^{-1}\, \int \md^d x\, \sqrt{g}\, \SR(g)$ from which $\eta_{\text{N}}$ was obtained as the scale derivative of the running Newton constant: $\eta_{\text{N}}=k\partial_k \ln G_k$.
Since in this term the metric $g_{\mu\nu}$ is to be interpreted as $g_{\mu\nu}=\bg_{\mu\nu}+\flcb_{\mu\nu}$, the running Newton constant fixes the normalization of the fluctuation field, $\flcb_{\mu\nu}$.
 While extremely tiny in magnitude, $\eta_{\text{N}}$ turned out negative with this entire class of truncations even in the `classical regime' displayed by the special (Einstein-Hilbert truncated, Type \Rmnum{3}a) trajectory which matches the observed values of Newton's constant and the cosmological constant \cite{frank1, h3, entropy}.

To see why the sign of the anomalous dimension is important let us consider an arbitrary field in $d$ spacetime dimensions with an inverse propagator $\propto Z(k^2) p^2$ which depends on an RG scale $k$. 
In absence of other relevant scales we may identify $k^2=p^2$, obtaining the dressed propagator $\widetilde{\mathcal{G}}(p) \propto \big[Z(p^2) p^2\big]^{-1}$. 
For example in a regime where $Z(k^2)\propto k^{-\eta}$ with a constant exponent $\eta$ we have, in momentum space, $\widetilde{\mathcal{G}}(p) \propto 1 \slash (p^2)^{1-\eta\slash2}$. If this propagator pertains to an {Euclidean} field theory on flat space it is natural to perform a Fourier transformation with respect to all $d$ coordinates, whence
\begin{align}
 \mathcal{G}_{\text{E}}(x-y) \propto \frac{1}{|x-y|^{d+\eta -2}} \label{eqn:anDim_01}
\end{align}
For field theories on Minkowski space the static limit of the propagator is particularly interesting; setting the time component of $p_{\mu}$ to zero and taking the $(d-1)$ dimensional Fourier transform of $\widetilde{\mathcal{G}}(p)$ we get, with $x \equiv (x^0,\vec{x})$ and $y\equiv (x^0,\vec{y})$ at equal times, 
\begin{align}
 \mathcal{G}_{\text{M}}(0,\vec{x}-\vec{y})\propto \frac{1}{|\vec{x}-\vec{y}|^{d+\eta-3}} \label{eqn:anDim_02}
\end{align}
Eqs. \eqref{eqn:anDim_01} and \eqref{eqn:anDim_02} confirm that the exponent $\eta$ which comes into play via the scale dependent field normalization $Z(k^2) \propto k^{-\eta}$ indeed deserves the name of an `anomalous dimension': the renormalization effects changed the effective dimensionality of spacetime, which manifests itself by the fall-off behavior of the 2-point function, from $d$ to $d+\eta$. In $d=3+1$, for instance, we obtain the modified Coulomb potential 
\begin{align}
 \mathcal{G}_{\text{M}}(0,\vec{x}-\vec{y}) \propto \frac{1}{|\vec{x}-\vec{y}|^{1+\eta}} \label{eqn:anDim_03}
\end{align}
The point to be noted here is that, {\it as compared to the classical Coulomb Green's function, a positive value of the anomalous dimensions renders the propagator more short ranged, while it becomes more long ranged when $\eta$ is negative.}

Thus we conclude that  the anomalous dimension $\eta_{\text{N}}<0$  found by the single-metric truncations of QEG corresponds to a graviton propagator on flat space which falls off for increasing distance {\it more slowly} than $1\slash |\vec{x}|$.
Also notice that, strictly speaking, eq. \eqref{eqn:anDim_01} holds only when $d+\eta-2 \neq 0$. If $d+\eta-2=0$ one has an increasing behavior even, $\mathcal{G}_{\text{E}}(x-y)\propto \ln (x-y)^2$. This is precisely the case relevant at the NGFP of quantum gravity where $\eta^*_{\text{N}}=-(d-2)$.
In the fixed point regime the momentum dependence is $\widetilde{\mathcal{G}}(p)\propto 1\slash p^d$. Note that at the NGFP  the function \eqref{eqn:anDim_02} becomes {\it linear:} $\mathcal{G}_{\text{M}}(0,\vec{x}-\vec{y})\propto  |\vec{x}-\vec{y}|$. 

The fall-off properties of the propagator have occasionally been adduced as a difficulty for the Asymptotic Safety idea.
We emphasize that in reality there is no such difficulty. 
It is nevertheless instructive to go through the argument, and to see where it fails.
For this purpose, consider an arbitrary bosonic quantum field $\Phi$ on $4D$ Minkowski space. Under very weak conditions one can derive a K\"{a}ll\'{e}n-Lehmann spectral representation \cite{Ka-Leh} for its dressed propagator:
\begin{align}
 \Delta_{\text{F}}^{\prime}(x-y)= \int_{0}^{\infty}\md \mu^2 \, \rho(\mu^2)\, \, \Delta_{\text{F}}(x-y;\mu^2) \label{eqn:anDim_04}
\end{align}
Here
\begin{align}
 \Delta_{\text{F}}(x-y;\mu^2)= - \int \frac{\md^4 p}{(2\pi)^4}\, \frac{e^{\Ii p (x-y)}}{p^2-\mu^2+\Ii \epsilon}
\end{align}
is the free Feynman propagator (with possible tensorial structures suppressed), and the spectral weight function
\begin{align}
 \rho(q^2)= (2\pi)^3 \sum_{\alpha} \delta^4(p_{\alpha}-q) \, | \langle 0 | \Phi(0) | \alpha\rangle |^2 \label{eqn:anDim_06}
\end{align}
contains a sum over all states $|\alpha \rangle$ with momenta $p_{\alpha}$ where $p_{\alpha}^2 \geq 0$, $p_{\alpha\, 0}\geq 0$ (the one-particle contribution included).
It is assumed that the states are elements of a vector space which is equipped with a positive-definite inner product.
Therefore it follows directly from its definition \eqref{eqn:anDim_06} that $\rho(\mu^2)$ is  a {\it non-negative} function. 
The K\"{a}ll\'{e}n-Lehmann representation itself follows from only a few, very basic additional assumptions: (a) completeness of the momentum eigenstates, in particular completeness of the asymptotic states, (b) the spectral condition $p^2\geq0$, $p_0\geq0$ for the states, (c) Poincar\'{e} covariance, in particular invariance of the vacuum state.

If a dressed propagator $\Delta_{\text{F}}^{\prime}$ possess a K\"{a}ll\'{e}n-Lehmann representation it follows that its Fourier transform behaves as $1\slash p^2$ for $p^2\rightarrow \infty$ limit, exactly as for the free one, $\Delta_{\text{F}}$. 
Conversely, for $|\vec{x}-\vec{y}|\rightarrow \infty$ at equal times, $\Delta_{\text{F}}^{\prime}$ cannot decay more slowly than $\propto 1 \slash |\vec{x}-\vec{y}|$. Indeed, the free massive Feynman propagator behaves as $\Delta_{\text{F}}(0,\vec{x}-\vec{y};\mu^2)\propto \exp\big( -\mu \,|\vec{x}-\vec{y}|\big)$ in this limit, so that the $\mu^2$-integral in \eqref{eqn:anDim_04} amounts to a superposition of decaying exponentials with non-negative weight, since $\rho(\mu^2)\geq0$. The best that can happen is that $\rho(\mu^2)$ has support at $\mu^2=0$, in which case the free propagator behaves Coulomb-like $\propto 1 \slash |\vec{x}-\vec{y}|$, and, as  a consequence, the dressed one as well, 
$\Delta_{\text{F}}^{\prime}(0,\vec{x}-\vec{y})\propto 1 \slash |\vec{x}-\vec{y}|$.
Obviously this is the behavior corresponding to an anomalous dimension $\eta=0$. If a K\"{a}ll\'{e}n-Lehmann representation exists, $\Delta_{\text{F}}^{\prime}$ may fall off faster, so $\eta>0$ is possible, but not more slowly.

As a consequence, under the conditions implying the existence of a K\"{a}ll\'{e}n-Lehmann representation negative anomalous dimensions $\eta<0$ cannot occur. This entails that, conversely, whenever an anomalous dimension is found to be negative one or several of those conditions must be violated. 

In the case of asymptotically safe gravity, described by the EAA, we can easily identify at least one of the above necessary conditions which is {\it not} satisfied:
The functional integral related to $\EAA_k[\flcb_{\mu\nu},\Ghx_{\mu},\GhAx^{\mu};\bg_{\mu\nu}]$ is a modified version (containing an IR regulator term) of the standard Faddeev-Popov gauge-fixed and BRST invariant functional integral which quantizes $\flcb_{\mu\nu}$ in some background gauge, usually the de Donder-Weyl gauge \cite{mr}.
However, the operatorial reformulation of this quantization scheme is well-known to involve a state space with an {\it indefinite metric} \cite{nakanishi-ojima}.
Therefore, $\rho(q^2)$ has no reason to be positive, and the short distance behavior of the dressed $\flcb_{\mu\nu}$ propagator may well be different from $1\slash p^2$ in momentum space.
In fact, Asymptotic Safety makes essential use of this possibility:
For $p^2\rightarrow \infty$, and in $d=4$, the propagator must be proportional to $1\slash p^4$ as a consequence of the UV fixed point.

A well-known example with similar properties is the Lorentz-covariant quantization of Yang-Mills theories on flat space, QCD, for instance. 
Here the anomalous dimension related to the gluon, $\eta\equiv \eta_{\text{F}}$, is negative too, and its negative sign is precisely the one responsible for asymptotic freedom.
Analogous to the computation done for the Newton constant, one can obtain $\eta_{\text{F}}$ in the EAA approach by using a (covariant) background type gauge and reading off $\eta_{\text{F}}$ from the term $\frac{1}{4 g_k^2}\int F_{\mu\nu}^2$ in $\EAA_k$ as the logarithmic scale derivative of the gauge coupling $g_k$, see ref. \cite{wett-mr} 
for details. 
A long ranged gluon propagator due to $\eta<0$ could be indicative of gluon confinement, at least in certain gauges.
Again the pertinent state space is not positive-definite, and so even propagators increasing with distance are not excluded by general principles.

It is actually quite intriguing that a linear confinement potential $\propto |\vec{x}-\vec{y}|$ for static color charges, corresponding to a $1\slash p^4$ behavior in the IR, is precisely what in gravity is realized in the UV.
While the fixed point regime of QEG is realized at small rather than large distances,  the graviton carries a {\it  large negative anomalous dimension} there.

Up to now we exploited only a rather technical, non-dynamical property of the quantization scheme used, namely the indefinite metric on state space, in order to reject the implications of a K\"{a}ll\'{e}n-Lehmann representation with a positive spectral density.
This was sufficient to demonstrate that within the setting of the (background gauge invariant) gravitational EAA of ref. \cite{mr} {\it the exact anomalous dimension derived from the running Newton constant is not bound to be positive for any general reason.}
Therefore there is nothing obviously wrong with the negative $\eta_{\text{N}}$'s that were found in concrete QEG calculations on truncated theory spaces, and a similar statement is true for Yang-Mills theory.

However, the previous argument has not yet much to do with the {\it dynamical} properties of the respective theory.
Taking QCD as an example again, we can solve the BRST cohomology problem which underlies its perturbative quantization,
and in this way we learn how to reduce the indefinite-metric state space to a subspace of  `physical' states which carries a positive definite inner product. 
One finds that, in this sense, transverse gluons and quarks are  `physical', while longitudinal and temporal gluons, as well as Faddeev-Popov ghosts are  `unphysical'.

Now, it is a highly non-trivial question whether the dynamics of the  `physical' states is such that the above requirements (a), (b), (c) are satisfied so that a K\"{a}ll\'{e}n-Lehmann representation of the transverse gluon propagator could exist.
The general believe is that the answer is negative since gluons, being confined, do not form a complete system of asymptotic states.
So here we have a deep dynamical rather than merely kinematical reason to reject the implications of the K\"{a}ll\'{e}n-Lehmann representation concerning the propagator's fall-off behavior.
This opens the door for a gluon propagator which might even increase with distance, like, for instance, the `IR enhanced' propagator proportional to $1\slash p^4$ for $p^2\rightarrow 0$.\footnote{Please note that by no means we are saying here that this behavior {\it must} occur, rather only that it {\it can} occur without violating any of the general principles discussed. In fact, detailed analyses of the IR properties of QCD, employing various independent non-perturbative techniques, indicate that in reality the picture is far more complex. For recent results concerning the gluon propagator, the properties of the spectral densities and the positivity properties of Yang-Mills theory we must refer to the literature \cite{realQCD}.}

Even though the gluon propagator is gauge dependent there is a direct connection to the gauge invariant confinement criterion of an area law for Wilson loops.
It has been shown \cite{west} that if the gluon propagator possesses the singular $1\slash p^4$ behavior for $p^2\rightarrow0$ in just {\it one} gauge then QCD is confining in the Wilson loop sense; in any other gauge it need not show this singular behavior.
In covariantly gauge fixed  QCD, it is of interest to know the properties of the gluon, ghost, and quark propagators also because they contain information about the nonperturbative dynamical mechanism by means of which the theory cuts down the indefinite state space to a positive-definite subspace, containing `physical' states only.

In gravity, the analogous question concerns the status of the transverse gravitons, that is, the $\flcb_{\mu\nu}$ modes which are not `pure gauge' but rather `physical' in the BRST sense.
Let us envisage a universe which, on all its vastly different scales, from the Planck regime to cosmological distances, is governed by QEG, and let us ask whether a transverse graviton which it may contain is more similar to a photon (unconfined, freely propagating, exists as an asymptotic state\footnote{to the extent this can make sense as an approximate notion in curved spacetime}) or to a gluon (confined, no asymptotic state, no K\"{a}ll\'{e}n-Lehmann representation with positive $\rho$)?

In its full generality this is a very hard question.
The attempt at an answer on the basis of existing single-metric computations would be that the graviton is more similar to the gluon than to the photon, a claim that might appear surprising, in particular if one thinks of astrophysical gravitational waves.

In quantum gravity, where Background Independence adds to the standard principles of quantum field theory, a particular convenient way to ensure a covariant formalism in presence of a gauge fixing and a cutoff term is the background field method. 
Thereby one introduces a generic background metric $\bar{g}_{\mu\nu}$ at intermediate stages of the quantization in addition to the usual dynamical metric $g_{\mu\nu}$.
Consequently, the most general ansatz for $\Gamma_k$ (possibly including matter fields) has to be of `bi-metric' type and thus contains all possible field monomials that can be constructed from $g_{\mu\nu}$, $\bar{g}_{\mu\nu}$, and the matter fields which respect all relevant symmetries (diffeomorphisms, (gauge-) symmetries in the matter sector, etc.)\cite{elisa2, MRS1}.
On the one hand, this affects for instance the mathematical description of the UV-completion of the theory, where a suitable UV fixed point defines the relation between the various invariants of background and physical metric.
However, on the other hand, the background is a purely technical artifice that does not affect the observables.
Those are obtained from the physical sector of $\Gamma_{k=0}\equiv \Gamma$ that mus respect Background Independence, i.e. which is independent of the auxiliary background field in the IR limit.
Hence, $\bar{g}_{\mu\nu}$ -- even though crucial in the construction of an exact RG flow at all intermediate scales $k$ -- becomes redundant when invoking fully intact split-symmetry in the IR ($k=0$). 
A technical simplification in the RG computations consists of neglecting the background invariants in the ansatz for the {\it non-gauge} part of the EAA, giving rise to the aforementioned single-metric approximation. Since the  gauge fixing and the cutoff action still rely explicitly on $\bar{g}_{\mu\nu}$, the RHS of the FRGE generates invariants depending on the background field and thus remain unresolved.
Hence, the reliability of the  single-metric approximation has to be tested by comparison of the class of solutions, in particular the UV completion with its more general bi-metric counterparts.
Thus, from a physical perspective, the ultimate theory (physical sector of $\Gamma_{k=0}$, on-shell S-matrix elements, for example) is {\it not} `bi-metric' in the sense that two independent metric tensors would play a role individually. 
There is only one physical metric; the background metric at intermediate stages of the quantization is only a technical artifice.
Fully intact split-symmetry in the physical sector, for vanishing IR cutoff, is precisely the statement that $\bar{g}_{\mu\nu}$ has become redundant and no observable depends on it.

The purpose of the present paper is to go beyond the single-metric approximation and investigate the crucial sign of the anomalous dimension $\eta_{\text{N}}$ using differently truncated functional RG flows of asymptotically safe metric gravity, i.e. QEG.
In particular we explore the corresponding predictions of two  `bi-metric truncations' of theory space \cite{elisa2, MRS1}. 
They have been studied recently in ref. \cite{MRS2}, henceforth denoted [\Rmnum{1}], and in ref. \cite{daniel2} which in the sequel is referred to as [\Rmnum{2}], respectively.
They employ a similar truncation ansatz for $\EAA_k[g,\bg]$, namely two {\it separate} Einstein-Hilbert terms for the dynamical and the background metric $g_{\mu\nu}$ and $\bg_{\mu\nu}$, respectively.
The calculations in [\Rmnum{1}] and [\Rmnum{2}] differ, however, with respect to the gauge fixing-conditions and -parameters they use, as well as the field parameterization they employ.
In [\Rmnum{1}] the `geometric' or `anharmonic' gauge fixing \cite{Mottola, anharmRoberto, frankmach, MRS2} is used, with gauge fixing parameter $\alpha=0$, while [\Rmnum{2}] relies on the harmonic gauge and $\alpha=1$.
Furthermore, in [\Rmnum{1}], the functional flow equation and in particular its mode suppression operator $\mathcal{R}_k$ was formulated in terms of a transverse-traceless (TT) decomposed field basis for $\flcb_{\mu\nu}$, no such decomposition was necessary in [\Rmnum{2}]. 
It is to be expected  that these differences of the coarse graining schemes employed  should have only a minor impact on the  RG flow and leave its essential qualitative features unchanged.

The rest of this paper is organized as follows.
In Section 2 we present a detailed analysis of the two bi-metric calculations [\Rmnum{1}], [\Rmnum{2}] and a comparison of their respective RG flows with the well-known one based on the single-metric Einstein-Hilbert truncation.
We demonstrate that the former imply a positive anomalous dimension, hence a `photon-like' behavior of gravitons in the semi-classical regime.
There is no obvious physical reason or qualitative argument that would explain the sign flip of $\eta$ in going from the single- to the bi-metric truncation. Therefore, detailed quantitatively precise calculations are particularly important here.
 
Section 3 is devoted to metric fluctuations outside this regime.
Their precise propagation properties near, but close to the Planck scale remain unknown for the time being.
We argue that, in this range of covariant momenta, they behave as a form of gravitating, but non-propagating `dark matter'.
Possible implications for the early Universe are also discussed.
Section 4 contains a brief summary.

\section{Anomalous dimension in single- and\\ bi-metric truncations}
Our approach to the quantization of  gravity  assumes that the fundamental degrees of freedom mediating the gravitational interaction are carried by the spacetime metric.
It heavily relies upon the Effective Average Action (EAA), a $k$-dependent functional $\EAA_k[g_{\mu\nu},\bg_{\mu\nu},\Ghx^{\mu},\GhAx_{\mu}]$ which, in the case of QEG, depends on the dynamical metric $g_{\mu\nu}$, the background metric $\bg_{\mu\nu}$, and the diffeomorphism ghost $\Ghx^{\mu}$ and anti-ghost $\GhAx_{\mu}$, respectively.
We employ the background field method to deal with the key requirement of Background Independence, and are thus led to the task of quantizing the metric fluctuations $\flcb_{\mu\nu}\equiv g_{\mu\nu}-\bg_{\mu\nu}$ in all  fixed but arbitrary backgrounds simultaneously.%
\footnote{In this paper we are dealing with pure gravity.
If one includes matter fields a general truncation ansatz for $\Gamma_k$ contains all possible field monomials that can be constructed from $g_{\mu\nu}$, $\bar{g}_{\mu\nu}$,  and the matter fields that respect the full set of imposed symmetries.
At the fundamental level ($k\rightarrow\infty$) the fixed point condition will fix the precise combination in which $g_{\mu\nu}$ and $\bar{g}_{\mu\nu}$ occur; in the final theory ($k\rightarrow 0$) instead it is, again, split-symmetry that forces one of the two metrics to become irrelevant or more precisely, `invisible' by the physical observables.}
 
For all truncations of theory space studied in this paper the corresponding ansatz for the EAA has the same general structure, namely
\begin{align}
 \EAA_k[g,\bg,\Ghx,\GhAx]&= \EAA^{\text{grav}}_k[g,\bg]+\EAA_k^{\text{gf}}[g,\bg]+ \EAA^{\text{gh}}_k[g,\bg,\Ghx,\GhAx] \label{eqn:trA02}
\end{align}
Concretely we consider the Einstein-Hilbert truncation, both  in its familiar single-metric form \cite{mr, frank1} and a more advanced bi-metric variant thereof \cite{MRS2,daniel2}. 
In the single-metric truncation the gravitational (`grav') part of the ansatz has the form
\begin{align}
  \EAA_k^{\text{grav}}[g,\bg]&= - \frac{1}{16\pi G_k^{\sm} } \int\md^d x \sqrt{g}\, \Big(\SR(g) - 2 \Kkbar^{\sm}\Big) \label{eqn:trA0sm}
\end{align}
It contains two running coupling constants, Newton's constant $G_k^{\sm}$ and the cosmological constant $\Kkbar_k^{\sm}$. (The superscript '$\sm$' stands for single-metric.)

For the most general bi-metric refinement of this truncation one should  in principle include the infinitely many invariants which one can construct from the metrics $g_{\mu\nu}$ and $\bg_{\mu\nu}$ that reduce to \eqref{eqn:trA0sm} when both metrics are identified, $g=\bg$.
Here, we follow earlier work in refs. \cite{MRS2} and \cite{daniel2}, from now on referred to as [\Rmnum{1}], [\Rmnum{2}], respectively, and retain for technical simplicity only four such invariants, namely two independent Einstein-Hilbert actions for $g$ and $\bg$, respectively:
\begin{align}
  \EAA_k^{\text{grav}}[g,\bg]&= - \frac{1}{16\pi \nkD } \int\md^d x \sqrt{g}\, \left(\SR(g) - 2 \KkbarD\right)\nonumber  \\
&\quad   - \frac{1}{16\pi \nkbB }  \int\md^d x \sqrt{\bg}\, \left(\SR(\bg)- 2 \KkbarB\right) \label{eqn:trA07}
\end{align}
This family of actions comprises 4 running coupling constants, the dynamical (`$\dyn$') Newton and cosmological constants as well as their background (`$\background$') counterparts.

An equivalent and sometimes more useful description of the action \eqref{eqn:trA07} is obtained by expanding $\EAA^{\text{grav}}_k[g,\bg]$ in powers of the fluctuation field $\flcb_{\mu\nu}=g_{\mu\nu}-\bg_{\mu\nu}$. 
We have, up to terms of second order in $\flcb_{\mu\nu}$:
\begin{align}
  \EAA_k^{\text{grav}}[\flcb;\bg]&= - \frac{1}{16\pi G_k^{(0)} }  \int\md^d x \sqrt{\bg} \left(\SR(\bg) - 2\Kkbar_k^{(0)}\right)  \nonumber  \\
&\quad - \frac{1}{16\pi G_k^{(1)} }  \int\md^d x \sqrt{\bg}\, \Big[-\bar{G}^{\mu\nu}-\Kkbar_k^{(1)} \bg^{\mu\nu}\Big] \flcb_{\mu\nu} \nonumber \\
&\quad - \frac{1}{2}  \int\md^d x \sqrt{\bg}\ \,\flcb^{\mu\nu} \,\,\EAA^{\text{grav}\,(2)}_k[\bg,\bg]\,\, \flcb_{\rho\sigma}
+ \Order{\flcb^3} \label{eqn:trA08}
\end{align}
This expansion in powers of $\flcb_{\mu\nu}$ is referred to as the `level representation' of the EAA, and a term is said to belong to level-($p$) if it contains $p$ factors of $\flcb_{\mu\nu}$, for $p=0,1,2,\cdots$.
The  level-($p$) couplings $G_k^{(p)},\,\Kkbar_k^{(p)}$, by definition, correspond to invariants that are of  order $(\flcb_{\mu\nu})^p$.  Their relation to the `$\dyn$' and `$\background$' couplings that were used in eq. \eqref{eqn:trA07} is given by, for $p=0$,
\begin{align}
 \frac{1}{G_k^{(0)}}=\frac{1}{G_k^{\background}}+\frac{1}{G_k^{\dyn}}\,,\qquad \quad 
\frac{\Kkbar_k^{(0)}}{G_k^{(0)}}=\frac{\Kkbar_k^{\background}}{G_k^{\background}}+\frac{\Kkbar_k^{\dyn}}{G_k^{\dyn}}\, ,
\label{eqn:trA08B}
\end{align}
and $G_k^{(p)}=G_k^{\dyn}$, $\Kkbar_k^{(p)}=\Kkbar_k^{\dyn}$ at all higher levels $p\geq1$.

Note that the couplings at level-(1) are precisely those which enter the field equation for self-consistent backgrounds, $\delta \EAA_k\slash \delta \flcb_{\mu\nu}|_{\flcb=0}=0$, while those at level-(2) and levels-($3,4,\cdots$) determine the propagator and the vertices of the $\flcb_{\mu\nu}$-self-interactions, respectively.
In the present truncation the latter roles are played by the same coupling namely $G_k^{(1)}=G_k^{(2)}=\cdots\equiv G_k^{\dyn}$, and likewise $\Kkbar_k^{(1)}=\Kkbar_k^{(2)}=\cdots\equiv \Kkbar_k^{\dyn}$.
However, it goes beyond a single-metric truncation as it resolves the differences between level-(0) and level-(1).\footnote{For structurally different calculations disentangling background and fluctuation fields see \cite{donkin-paw,paw-rodigast,codello-closure,matterPerc,morris-dietz}.}
Single-metric calculations retain only terms of order $(\flcb_{\mu\nu})^0$, i.e. of level-(0), and then postulate that the RG running of the couplings at the higher levels is well approximated by that at level-(0). (See \cite{daniel2} for a detailed discussion.)

The gauge fixing and the ghost terms $\EAA_k^{\text{gf}}$ and $\EAA_k^{\text{gh}}$ in \eqref{eqn:trA02} are determined by the gauge fixing function
\begin{align}
 \mathcal{F}_{\mu}^{\alpha\beta}[\bg] \, \flcb_{\mu\nu}&\equiv \big(\delta^{\beta}_{\mu}\bg^{\alpha\gamma} \bZ_{\gamma}-\varpi \bg^{\alpha\beta}\bZ_{\mu}\big)\,\flcb_{\mu\nu} 	\label{eqn:trA03}
\end{align}
which involves a free parameter, $\varpi$, whose RG running is neglected here.
Special cases include the harmonic gauge ($\varpi=1\slash2$) and the geometric, or `anharmonic' gauge ($\varpi=1\slash d$). 
In addition there appears the  gauge parameter $\alpha$ in the gauge fixing action whose $k$-dependence will be neglected as well:
\begin{align}
\EAA_k^{\text{gf}}[g,\bg]&= \frac{1}{32\pi \alpha \,G^{\dyn\slash \sm}_k} \int\md^d x \sqrt{\bg} \,\,\bg^{\mu\nu} \Big[\mathcal{F}_{\mu}^{\alpha\beta}[\bg] \left(g_{\alpha\beta}-\bg_{\alpha\beta}\right) \Big] \Big[\mathcal{F}_{\nu}^{\rho\sigma}[\bg] \left(g_{\rho\sigma}-\bg_{\rho\sigma}\right) \Big]		\label{eqn:trA04}
\end{align}
Specifically, the two gauge fixing parameters were chosen as $(\varpi=1\slash2,\,\alpha=1)$, $(\varpi=1\slash d,\, \alpha\rightarrow0$), and $(\varpi=1\slash2,\, \alpha=1)$ in the single-metric truncation of \cite{mr}, the `TT-decomposed'\footnote{The Hessian of $\EAA_k$ in the Einstein-Hilbert truncation contains uncontracted derivative operators such as $\bZ_{\mu}\bZ_{\nu}$. In  [\Rmnum{1}] a transverse-traceless (TT) decomposition of the fluctuation field $\flcb_{\mu\nu}$ was employed to deal with this complication. The problematic operators act on the component fields as fully contracted Laplacian $\bg^{\mu\nu}\bZ_{\mu}\bZ_{\nu}$ then, and  heat kernel methods can be applied to evaluate the functional traces due to the various irreducible fields.} bi-metric calculation of [\Rmnum{1}], and the `$\cf$-deformed'\footnote{In [\Rmnum{2}], $\cf$ denotes a conformal parameter introduced as a tool to distinguish between dynamical and background contributions. The freedom in choosing a gauge parameter $\alpha$ was exploited to reduce the functional trace on the RHS of the FRGE to a function of the Laplacian $\bZ^2$ alone, which then could be computed using standard heat kernels again.} bi-metric analysis in [\Rmnum{2}], respectively.

When the full ansatz is inserted into the functional renormalization group equation (FRGE) we obtain a coupled system  of RG differential equations which, when expressed in terms of dimensionless couplings\footnote{The dimensionless couplings, $\tg^{\cix}_k$ and $\Kk_k^{\cix}$, are related to the dimensionful ones, $G_k^{\cix}$ and $\Kkbar_k^{\cix}$, appearing in the truncation ansatz, by  $G_k^{\cix}=k^{2-d}\tg_k^{\cix}$ and $\Kkbar_k^{\cix}=k^2 \Kk_k^{\cix}$, respectively.}, has the following structure:  
\begin{subequations}
\begin{align}
 &\partial_t \tg_k^{\dyn\slash \sm}
 = 
  \big[d-2+\eta^{\dyn\slash \sm}(\tg_k^{\dyn\slash \sm},\Kk^{\dyn\slash \sm}_k)\big]\tg_k^{\dyn\slash \sm}\label{eqn:trA27CA}\\
&\partial_t \Kk^{\dyn\slash \sm}_k= \beta_{\Kk}^{\dyn\slash \sm}(\tg_k^{\dyn\slash \sm},\Kk^{\dyn\slash \sm}_k)\label{eqn:trA27CB}\\
&\partial_t \tg_k^{(0)}= 
 \big[d-2+\eta^{(0)}(\tg_k^{\dyn},\KkD_k,\tg_k^{(0)})\big]\tg_k^{(0)}\label{eqn:trA27CC}\\
&\partial_t \Kk^{(0)}_k= \beta_{\Kk}^{(0)}(\tg_k^{\dyn},\KkD_k,\tg_k^{(0)},\Kk^{(0)}_k)\label{eqn:trA27CD}
\end{align}\label{eqn:trA27C}
\end{subequations}
The two  equations \eqref{eqn:trA27CA} and eq. \eqref{eqn:trA27CB} constitute the single-metric system, while the bi-metric system is described by the full set of all 4 differential equations.

Since the above equations are partially decoupled, solutions $k\mapsto \left(\tg_k^{\dyn},\KkD_k, \tg^{(0)}_k,\Kk^{(0)}_k\right)$ can be obtained from \eqref{eqn:trA27C} in a hierarchical fashion: $(\tg_k^{\dyn},\,\KkD_k)\Rightarrow \tg_k^{(0)}\Rightarrow \Kk^{(0)}_k$. 
Notice that the explicit form of the beta-functions to be used is different for the three truncations we are going to consider here; they can be found in \cite{mr}, [\Rmnum{1}], and [\Rmnum{2}], respectively.

In the sequel, we mostly focus on the Newton couplings $G_k^{\cix}$ and their non-canonical RG running which is described by the respective anomalous dimension $k\partial_k \ln G_k^{\cix}\equiv\eta^{\cix}$.
In all truncations considered here its general structure is
\begin{align}
\eta^{\cix}&= \frac{B_1^{\cix}(\Kk)\,\tg^{\cix}}{1-B_2^{\cix}(\Kk)\,\tg^{\cix}} \quad \text{ for } \cix\in\{\dyn,\,\background,\,{(0)},\,\sm\}
\label{eqn:res4D_005}
\end{align}
The level- and background-$\eta^{\cix}$'s are related by $\eta^{(0)}\slash g^{(0)}=\eta^{\background}\slash g^{\background}+\eta^{\dyn}\slash \tg^{\dyn}$.

In the sequel we employ the language of levels and always present the couplings of the $\flcb_{\mu\nu}$-independent invariants, denoted by a superscript (0), together with the higher level couplings which are collectively denoted by '$\dyn$', standing for $(p)$, $p\geq 1$. (The `$\background$' couplings could be obtained from \eqref{eqn:trA08B} if needed.)

In the following subsections we analyze the anomalous dimensions related to the various versions of Newton's constant. We begin with the single-metric case and then proceed to  the two   bi-metric calculations [\Rmnum{1}] and [\Rmnum{2}]. 

Unless stated otherwise, we always assume 4 spacetime dimensions ($d=4$) in the rest of this paper, and we employ the optimized cutoff shape function \cite{litimPRL}.

\subsection{Single-metric truncation}
In the single-metric Einstein-Hilbert truncation the RG running of Newton's constant is governed by
\begin{align}
\eta^{\sm}(\tg^{\sm},\Kk^{\sm})&= \frac{B_1^{\sm}(\Kk^{\sm})\,\tg^{\sm}}{1-B_2^{\sm}(\Kk^{\sm})\,\tg^{\sm}}
\label{eqn:res4D_005B}
\end{align}
\begin{figure}[h!]
\subfloat[The sign of $\eta^{\sm}$.]{
\psfrag{d}[bc][bc][1][90]{${\scriptstyle B_1^{\dyn}(\Kk;4)-B_1^{(0)}(\Kk;4)}$}
\psfrag{g}[bc][bc][1][270]{${\scriptstyle  \tg^{\sm}}$}
 \psfrag{l}{${\scriptstyle \Kk^{\sm}}$}
\psfrag{b}[bc][bc][1][90]{${\scriptstyle B_1^{\dyn\slash(0)}(\Kk;4)}$}
\centering
\includegraphics[width=0.49\textwidth]{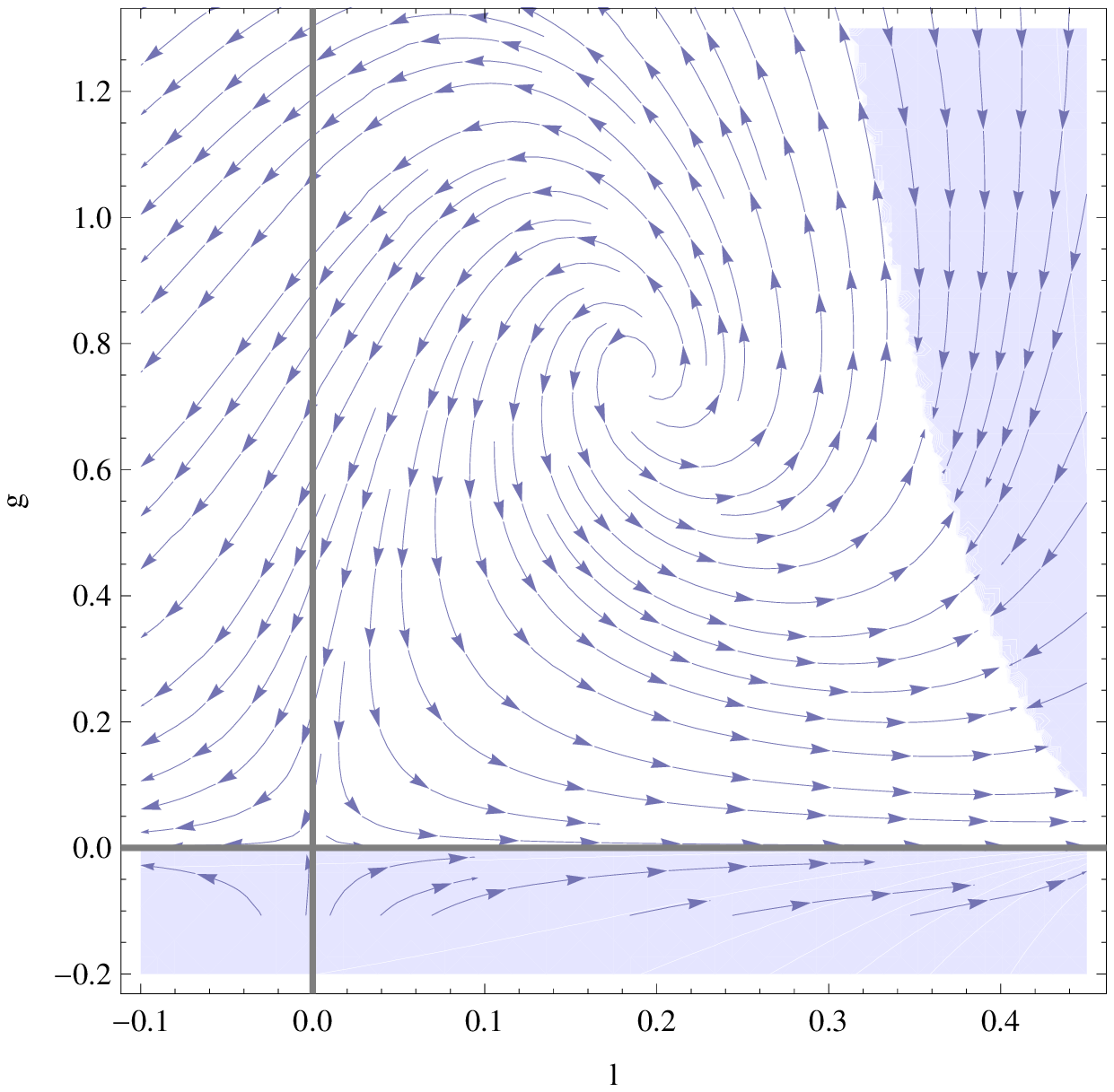}
 \label{fig:b1db1smSM}
}
\subfloat[The contour plot of $\eta^{\sm}$.]{
 \psfrag{g}[c]{${\scriptstyle  \tg^{\sm}}$}
  \psfrag{l}{${\scriptstyle \Kk^{\sm}}$}
 \label{fig:etaContourSM}\includegraphics[width=0.48\textwidth]{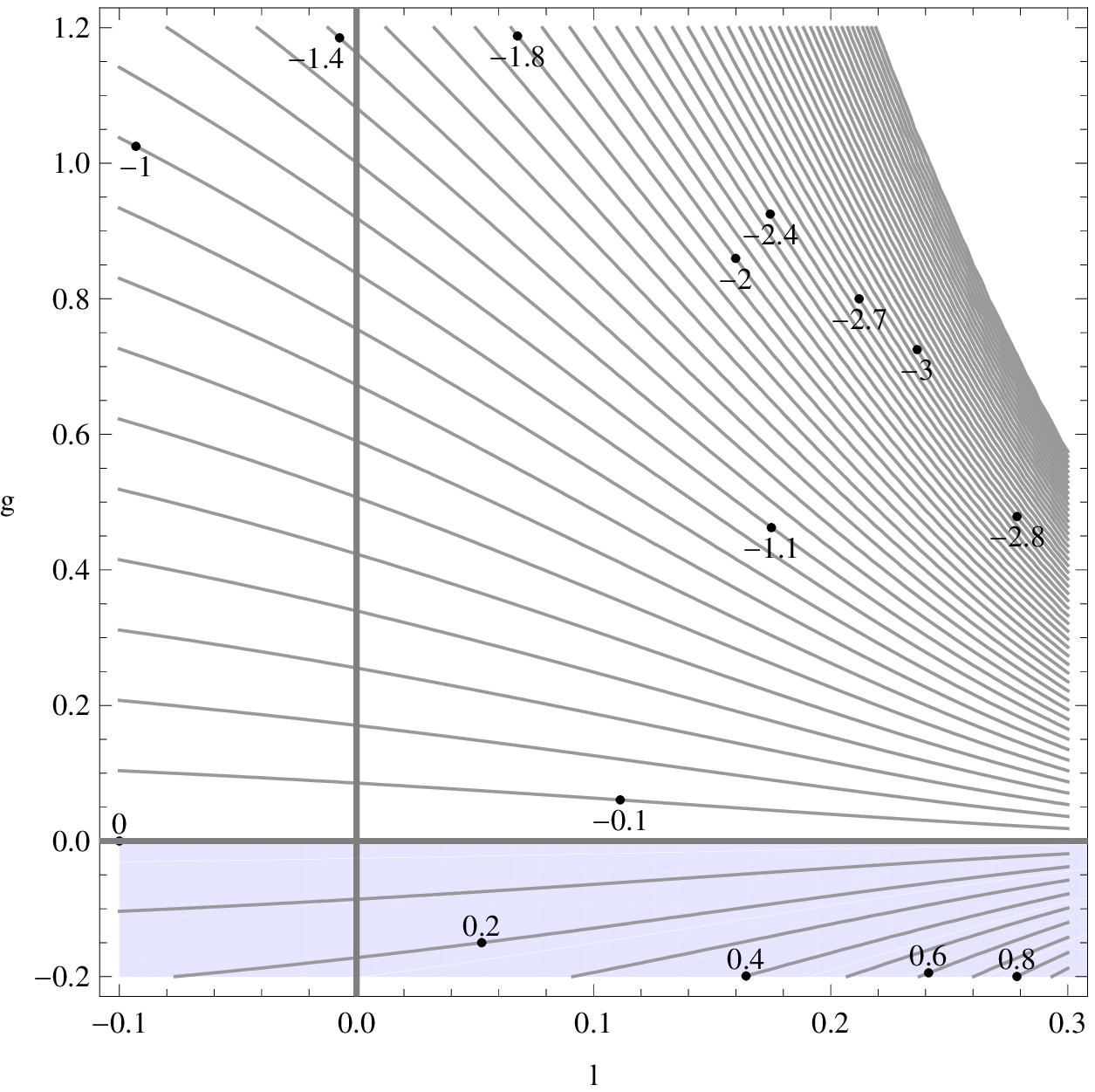}}
\caption{The phase-portrait of the single-metric Einstein-Hilbert truncation. The shaded areas in the left diagram indicate regions in the $(\tg^{\sm}, \Kk^{\sm})$-plane of positive anomalous dimension $\eta^{\sm}$. In the single-metric approximation, $\eta^{\sm}$ is seen to be negative everywhere on the physically accessible part of theory space.  The contour plot of the right diagram shows the lines of constant $\eta^{\sm}$ values (`iso-$\eta$' lines).} 
 \label{fig:etaSM}
\end{figure}
The function $B_1^{\sm}(\Kk^{\sm})$ in the numerator of eq. \eqref{eqn:res4D_005B} is given by 
\begin{align}
B_1^{\sm}(\Kk^{\sm})&=-\frac{1}{3\pi}\Big\{ 
-5\ThrfA{1}{1}{-2\Kk^{\sm}}+18 \ThrfA{2}{2}{-2\Kk^{\sm}}
+ 4\ThrfA{1}{1}{0}+6 \ThrfA{2}{2}{0} 
\Big\}
\label{eqn:qeg010}
\end{align}
and   $B_2^{\sm}(\Kk^{\sm})$ in the denominator reads
\begin{align}
B_2^{\sm}(\Kk^{\sm})&=\frac{1}{6\pi}\Big\{ 
-5\ThrfB{1}{1}{-2\Kk^{\sm}}+18 \ThrfB{2}{2}{-2\Kk^{\sm}}
\Big\}
\label{eqn:qeg011}
\end{align}
Here $\Phi$ and $\widetilde{\Phi}$ are the standard threshold functions introduced in \cite{mr} which depend on the details of the cutoff scheme, its `shape function' $R^{(0)}$ in particular.

We are interested in the sign of $\eta^{\sm}$ in dependence on $\tg^{\sm}$ and $\Kk^{\sm}$, the two coordinates on theory space.
As can be seen from the plot in Fig. \ref{fig:b1db1smSM}, in the single-metric truncation, {\it the anomalous dimension $\eta^{\sm}$ is negative in the entire physically relevant region of the $\tg^{\sm}$-$\Kk^{\sm}$ theory space.} 
This is a well-known fact, already mentioned in the Introduction, and has been confirmed also by all single-metric truncations with more than the $\int\sqrt{g} $ and $\int \sqrt{g}\SR$ terms in the ansatz that were analyzed so far \cite{oliver2,oliver3,anharmRoberto,frankmach,Saueressig:2011vn,BMS,BMS2,creh2,creh3,frank-fR,stefan-frankfrac,morris-dietz-fR,Benedetti-relevant,ohtaPerc}.

In the semi-classical regime\footnote{To be precise, we consider a `type \Rmnum{3}a' trajectory here, which, by definition, has a {\it positive} cosmological constant in the IR, see \cite{frank1}.} where $0<\tg^{\sm},\,\Kk^{\sm}\ll 1$ the term $B_2^{\sm}(\Kk^{\sm})\tg^{\sm}$ in the denominator on the RHS of \eqref{eqn:res4D_005B} is negligible, hence the negative sign of $\eta^{\sm}$ is entirely due to the negative sign of $B_1^{\sm}(\Kk^{\sm})$ that occurs for small arguments $\Kk^{\sm}\ll 1$. 
Here it is a reliable approximation to set $\eta^{\sm}\approx B_1^{\sm}(\Kk^{\sm})\tg^{\sm}$.

It is instructive to expand the function $B_1^{\sm}$ for small values of the (dimensionless) cosmological constant:
\begin{align}
B_1^{\sm}(\Kk^{\sm})&=\frac{1}{3\pi}\Big[ 
\ThrfA{1}{1}{0}-24 \ThrfA{2}{2}{0}
\Big]-\frac{26}{3\pi}\Kk^{\sm}+ \Order{(\Kk^{\sm})^2}
\label{eqn:qeg012}
\end{align}
This linear approximation  confirms the negative values of $B_1^{\sm}$  in the semi-classical regime:
Its $\Kk^{\sm}$-independent term $B_1^{\sm}(0)$ is known to be negative for any admissible cutoff \cite{mr}, and the term linear in the cosmological constant is negative, too, when $\Kk^{\sm}>0$. 

Notice that the slope of the linear function \eqref{eqn:qeg012} is {\it universal}, i.e. cutoff scheme independent.
Every choice of the shape function $R^{(0)}$ used in the threshold functions $\Phi$ and $\widetilde{\Phi}$ yields the same slope, $-26\slash 3\pi$, which is negative and thus favors an anomalous dimension which is negative, too. 
The constant term in \eqref{eqn:qeg012} is cutoff scheme dependent, however its negative sign is not. Hence, starting from $B_1^{\sm}(0)<0$, the function $B_1^{\sm}(\Kk^{\sm})$, and therefore also $\eta^{\sm}(\tg^{\sm},\Kk^{\sm})$, decreases with increasing values of $\Kk^{\sm}$, and in fact stays negative throughout the relevant part of theory space ($\Kk^{\sm}<1\slash 2$).

\subsection{The (TT-based) bi-metric calculation [\Rmnum{1}]}
Turning to truncations of bi-metric type now, let us consider the approach followed in [\Rmnum{1}] first.  
\begin{figure}[h!]
 \subfloat[The $\dyn$-sector.]{
 \psfrag{g}{${\scriptstyle  \tg^{\dyn}}$}
  \psfrag{l}{${\scriptstyle \KkD}$}
 \label{fig:etaMRS2}\includegraphics[width=0.4\textwidth]{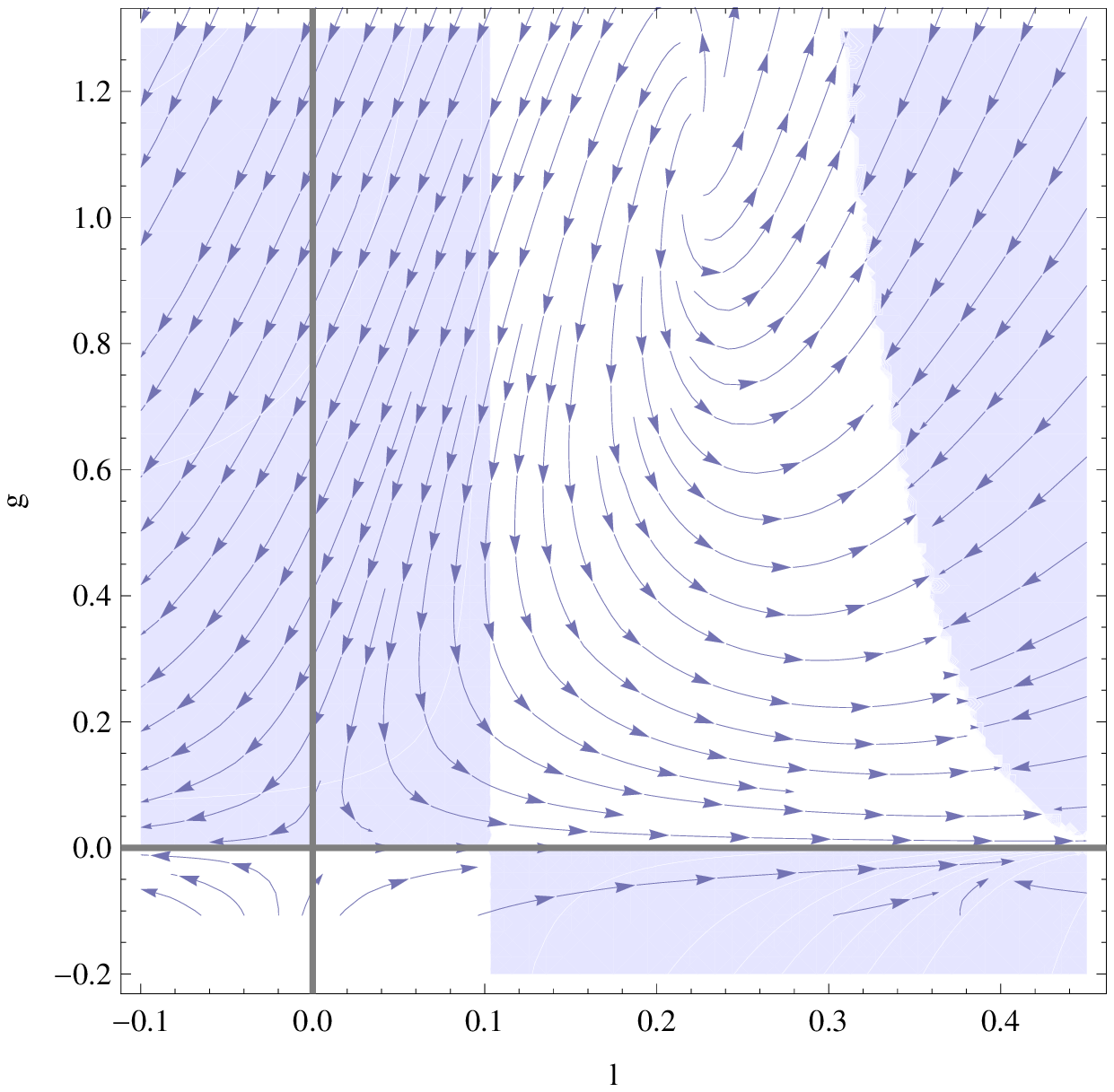}}
 \hspace{0.1\textwidth}
 \subfloat[The level-(0) sector.]{
 \psfrag{g}{${\scriptstyle  \tg^{(0)}}$}
  \psfrag{l}{${\scriptstyle \Kk^{(0)}}$}
 \label{fig:etaMRS20}\includegraphics[width=0.4\textwidth]{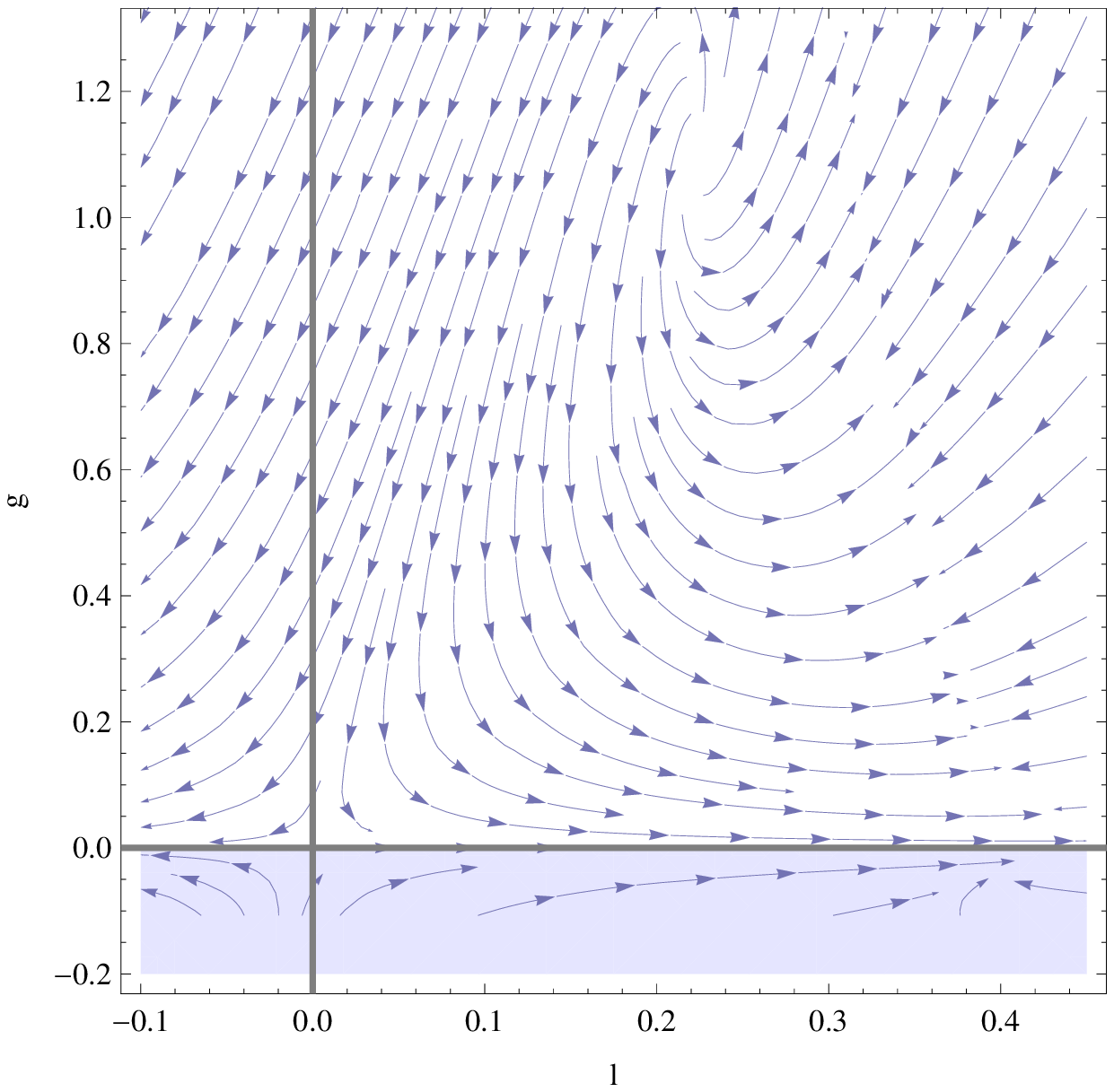}}
\caption{The phase portraits on the $(\tg^{\dyn},\KkD)$- and the $(\tg^{(0)},\Kk^{(0)})$-plane, respectively, obtained using the bi-metric results of [\Rmnum{1}]. The shaded (white) areas are regions of positive (negative) anomalous dimension $\eta^{\dyn}$ and $\eta^{(0)}$, respectively. While the dynamical anomalous dimension $\eta^{\dyn}$ exhibits regions of physical interest where it is  positive, $\eta^{(0)}$ does not.}  \label{fig:b1db1smMRS2LD}
\end{figure}
In the dynamical sector the dependence of the corresponding anomalous dimension 
\begin{align}
\eta^{\dyn}(\tg^{\dyn},\Kk^{\dyn})&= \frac{B_1^{\dyn}(\Kk^{\dyn})\,\tg^{\dyn}}{1-B_2^{\dyn}(\Kk^{\dyn})\,\tg^{\dyn}}
\label{eqn:res4D_005C}
\end{align} 
on the cosmological constant $\Kk^{\dyn}$ is described by the numerator function,
\begin{align}
B_1^{\dyn}(\KkD)&=\frac{1}{6\pi}\Big\{
25\,\ThrfA{2}{2}{-2\KkD}+ \ThrfA{2}{2}{-\tfrac{4}{3}\KkD}-80\, \ThrfA{3}{3}{-2\KkD}
\nonumber \\ & \qquad \qquad
-3\,\ThrfA{2}{2}{0}+28\,\ThrfA{3}{3}{0}+72\,\ThrfA{4}{4}{0}
\Big\}
\end{align}
while the denominator contribution in eq. \eqref{eqn:res4D_005C} contains
\begin{align}
B_2^{\dyn}(\KkD)&=-\frac{1}{12\pi}\Big\{
25\,\ThrfB{2}{2}{-2\KkD}+ \ThrfB{2}{2}{-\tfrac{4}{3}\KkD}-80\,\ThrfB{3}{3}{-2\KkD}
\Big\}
\end{align}

The beta-functions of the level-(0) and the background-sector are sensitive to the dynamical couplings as well. 
In particular the sign of the anomalous dimension $\eta^{(0)}$, pertaining to the {\it level-(0)} Newton constant $\tg^{(0)}$, is strongly dependent on the {\it dynamical} cosmological constant, $\KkD$. Explicitly,
\begin{align}
 \eta^{(0)}(\tg^{\dyn},\KkD,\tg^{(0)})&=
 \frac{1}{12\pi}\, \Big[10\, \qA{1}{1}{-2\KkD} + 2\, \qA{1}{1}{-\frac{4}{3}\KkD}
- 40\, \qA{2}{2}{-2\KkD} \nonumber \\
&\quad\qquad \quad  -   8 \, \ThrfA{1}{1}{0}   - 13\,\ThrfA{2}{2}{0}\Big]\tg^{(0)}
\label{eqn:res4D_003}
\end{align}
where $\qA{p}{n}{w}\equiv \ThrfA{p}{n}{w}-\tfrac{\eta^{\dyn}}{2}\ThrfB{p}{n}{w}$ involves $\eta^{\dyn}\equiv\eta^{\dyn}(\tg^{\dyn},\KkD)$.

In Figs. \ref{fig:etaMRS2} and \ref{fig:etaMRS20} we display the $(\tg^{\dyn},\KkD)$ and $(\tg^{(0)},\Kk^{(0)})$ phase portraits for the dynamical and the level-(0) couplings, respectively. 
For the latter, the overall picture is essentially the same as for the single-metric truncations: 
The anomalous dimension $\eta^{(0)}$ is negative everywhere on theory space (where $\tg^{(0)}>0$), in particular in the semi-classical regime.
\begin{figure}[ht!]
\centering
 \psfrag{c}[rb]{${\scriptstyle  \KkD_{\crit}}$}
 \psfrag{g}[c]{${\scriptstyle  \tg^{\dyn}}$}
  \psfrag{l}{${\scriptstyle \Kk^{\dyn}}$}
 \includegraphics[width=0.5\textwidth]{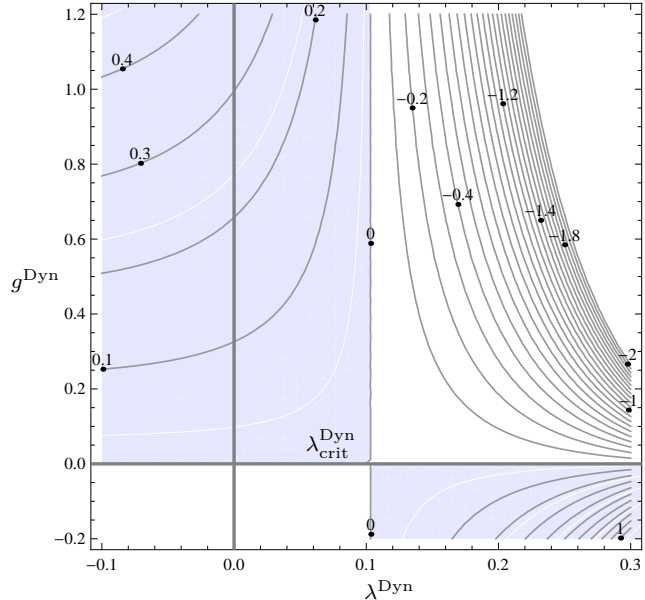}
\caption{The contour plot shows the lines of constant anomalous dimension $\eta^{\dyn}$ implied by the bi-metric calculation [\Rmnum{1}]. The shaded regions correspond to $\eta^{\dyn}>0$. Note the sign flip of $\eta^{\dyn}$ at a non-zero critical value of the cosmological constant, $\KkD_{\crit}>0$.}   \label{fig:etaContourMRS2}
\end{figure}
However, the dynamical $(\tg^{\dyn},\KkD)$-flow reveals a novel aspect of the bi-metric truncation: the anomalous dimension $\eta^{\dyn}$ is {\it positive} for $\KkD$ smaller than a certain critical value $\KkD_{\crit}>0$, and turns negative only when $\KkD>\KkD_{\crit}$.

While this conclusion is drawn on the basis of the complete formula \eqref{eqn:res4D_005} including the denominator, the sign of $\eta^{\dyn}$ coincides with the sign of $B_1^{\dyn}$ since $B_2^{\dyn}\tg^{\dyn}\ll 1$ in the entire region of interest, so that $\eta^{\dyn}\approx B_1^{\dyn}(\KkD)\tg^{\dyn}$ is a good approximation.
As a consequence, the domains where $\eta^{\dyn}>0$ and $\eta^{\dyn}<0$, respectively, are separated by a straight line on the $(\tg^{\dyn},\KkD)$-plane located at $\KkD=\KkD_{\crit}$ with $B_1^{\dyn}(\KkD_{\crit})=0$.

The sign flip of $B_1^{\dyn}$  can be  demonstrated  analytically by expanding $B_1^{\dyn}$ in powers of the cosmological constant:
\begin{align}
B_1^{\dyn}(\Kk^{\dyn})&=\frac{1}{6\pi}\Big[ 
23\ThrfA{2}{2}{0}+72\ThrfA{4}{4}{0}-72\ThrfA{3}{3}{0}
\Big]-\frac{43}{9\pi}\KkD+ \Order{(\Kk^{\dyn})^2}
\end{align}
Again, the slope of this linear function is found to be both universal and negative, $-43\slash 9\pi$ in this case. 
The difference in comparison with the single-metric truncation lies in the constant term $B_1^{\dyn}(0)$: according to the bi-metric calculation it is  {\it positive} for all plausible cutoff schemes, in sharp contradistinction to $B_1^{\sm}(0)<0$ in the single-metric case.

For the example of the optimized shape function we have $B_1^{\dyn}(0)=+35\slash 36\pi$, yielding the critical cosmological constant  $\KkD_{\crit}|_{\text{opt.cutoff}}\approx 0.2$.  The quadratic terms $\propto (\KkD)^2$ in $B_1^{\dyn}$ correct this result to about $\KkD_{\crit}|_{\text{opt.cutoff}}\approx 0.1$ which is then  stable under the addition of still higher orders and coincides with the exact value.

So we may conclude that  {\it the dynamical anomalous dimension $\eta^{\dyn}$ is positive in the semi-classical regime where $0<\tg^{\dyn},\,\KkD\ll 1$} and becomes negative for $\KkD>\KkD_{\crit}\approx 0.1$.
This is also seen in Fig. \ref{fig:etaContourMRS2} which shows the lines of constant $\eta^{\dyn}$ values on the $(\tg^{\dyn},\KkD)$ plane. 
Recall that the NGFP, for instance, is located on the curve with $\eta^{\dyn}=-2$.
\begin{figure}[h!]
\psfrag{b}[bc]{${\scriptstyle B^{\dyn}_{1}(0)}$}
\psfrag{a}[bc]{${\scriptstyle s}$}
\psfrag{t}[tc]{${\scriptstyle }$}
\psfrag{r}[bc]{${\scriptstyle }$}
\centering
\includegraphics[width=0.6\textwidth]{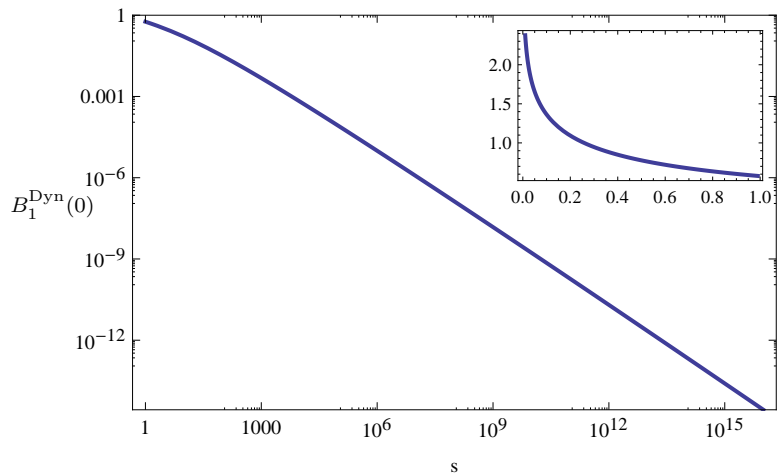}
\caption{The constant $ B^{\dyn}_{1}(0)$ as obtained in [\Rmnum{1}] for different values of the shape parameter $s$ characterizing the exponential shape functions. While cutoff scheme dependent, its sign is always positive, implying the existence of a critical cosmological constant $\KkD_{\crit}>0$ such that $\eta^{\dyn}>0$ for $\KkD< \KkD_{\crit}$.}  \label{fig:b1db1MRS2}
\end{figure}

To verify that the novel feature of a positive $\eta^{\dyn}$  in the semi-classical regime is independent of the cutoff-scheme chosen,  we have checked the corresponding  condition $B_1^{\dyn}(0)>0$ for the one-parameter family of exponential shape functions $\,R^{(0)}(y)=sy\,\left[\exp(s y)-1\right]^{-1}$, for example \cite{souma, frank1, oliver1, oliver2}.
Its threshold functions at vanishing argument can be evaluated exactly:
\begin{align}
B_1^{\dyn}(0)&=\frac{1}{6\pi (s-1)^3}\Big[ 
-(7s-3)(s-1)+\big(9+s(14s-19)\big)\ln(s)
\Big]
\end{align}
This result for the $s$-dependence, plotted in Fig. \ref{fig:b1db1MRS2}, is indeed reassuring: even though the value of $B_1^{\dyn}(0)$ decreases for increasing `shape parameter' $s$, it stays always positive.
This yields a critical value $\KkD_{\crit}$ which is positive, too.
Thus, the bi-metric calculation [\Rmnum{1}], very robustly, predicts a semi-classical regime with a positive value of the dynamical anomalous dimension $\eta^{\dyn}$.

\subsection{The (`$\bm{\cf}$-deformed') bi-metric calculation [\Rmnum{2}]}
A different bi-metric approach that is more closely related to the single-metric computation in \cite{mr} was developed in  [\Rmnum{2}] recently. 
While it employs the same truncation ansatz, namely two separate Einstein-Hilbert actions for $g_{\mu\nu}$ and $\bg_{\mu\nu}$, the gauge fixing and the field parametrization chosen are different from the calculation [\Rmnum{1}].
In order to explore whether the novel properties displayed by [\Rmnum{1}] are actually due to its bi-metric character, and to what extent gauge fixing and field parametrization issues play a role possibly, we shall now repeat the analysis of the previous subsection, this time using the beta-functions obtained in [\Rmnum{2}]. 
\begin{figure}[h!]
\subfloat[The $\dyn$-sector.]{
 \psfrag{g}{${\scriptstyle  \tg^{\dyn}}$}
  \psfrag{l}{${\scriptstyle \KkD}$}
 \label{fig:b1db1smBMLD}\includegraphics[width=0.4\textwidth]{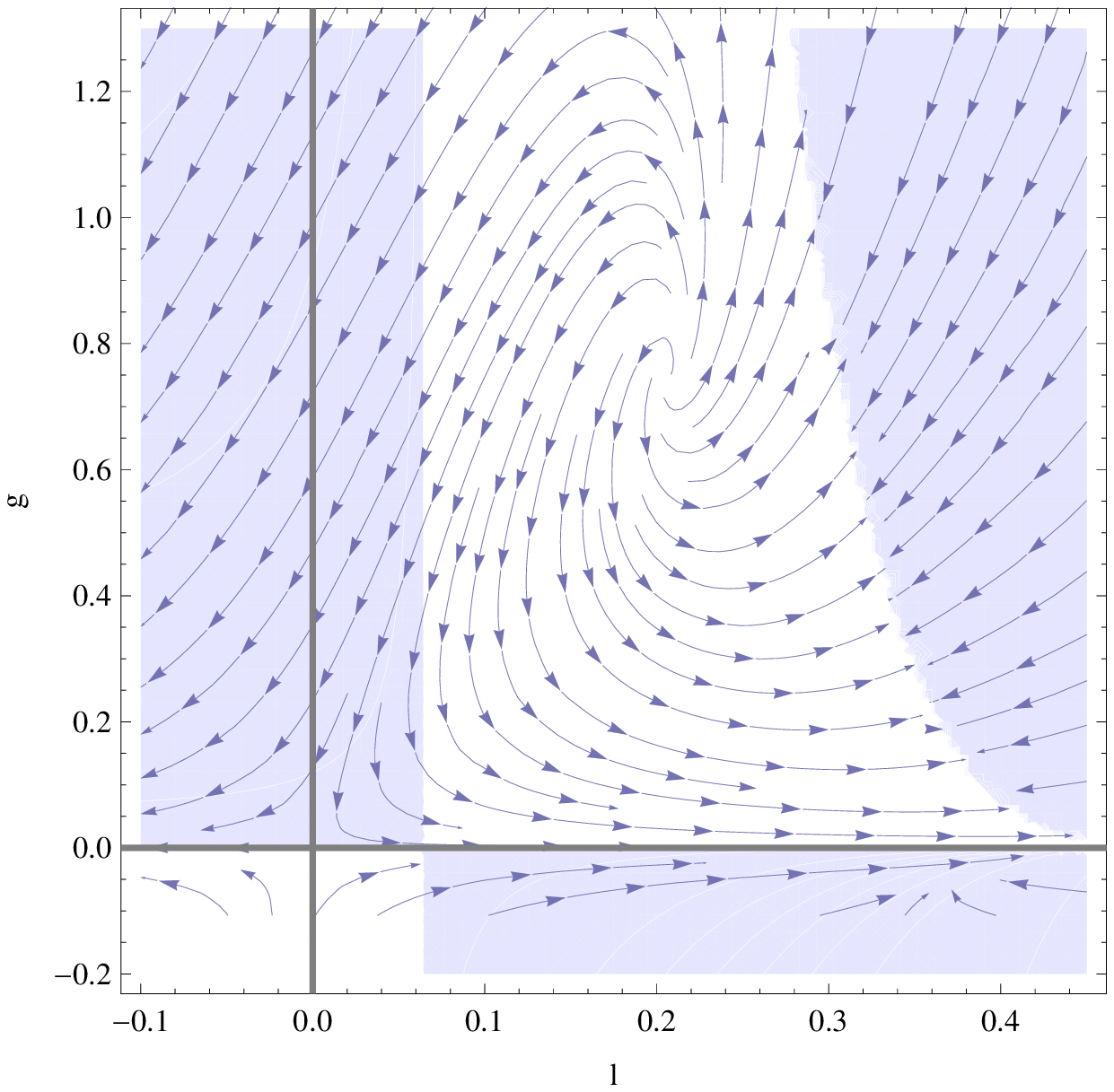}}
 \hspace{0.1\textwidth}
 \subfloat[The level-(0) sector.]{
 \psfrag{g}{${\scriptstyle  \tg^{(0)}}$}
  \psfrag{l}{${\scriptstyle \Kk^{(0)}}$}
  \label{fig:b1db1smBML0}\includegraphics[width=0.4\textwidth]{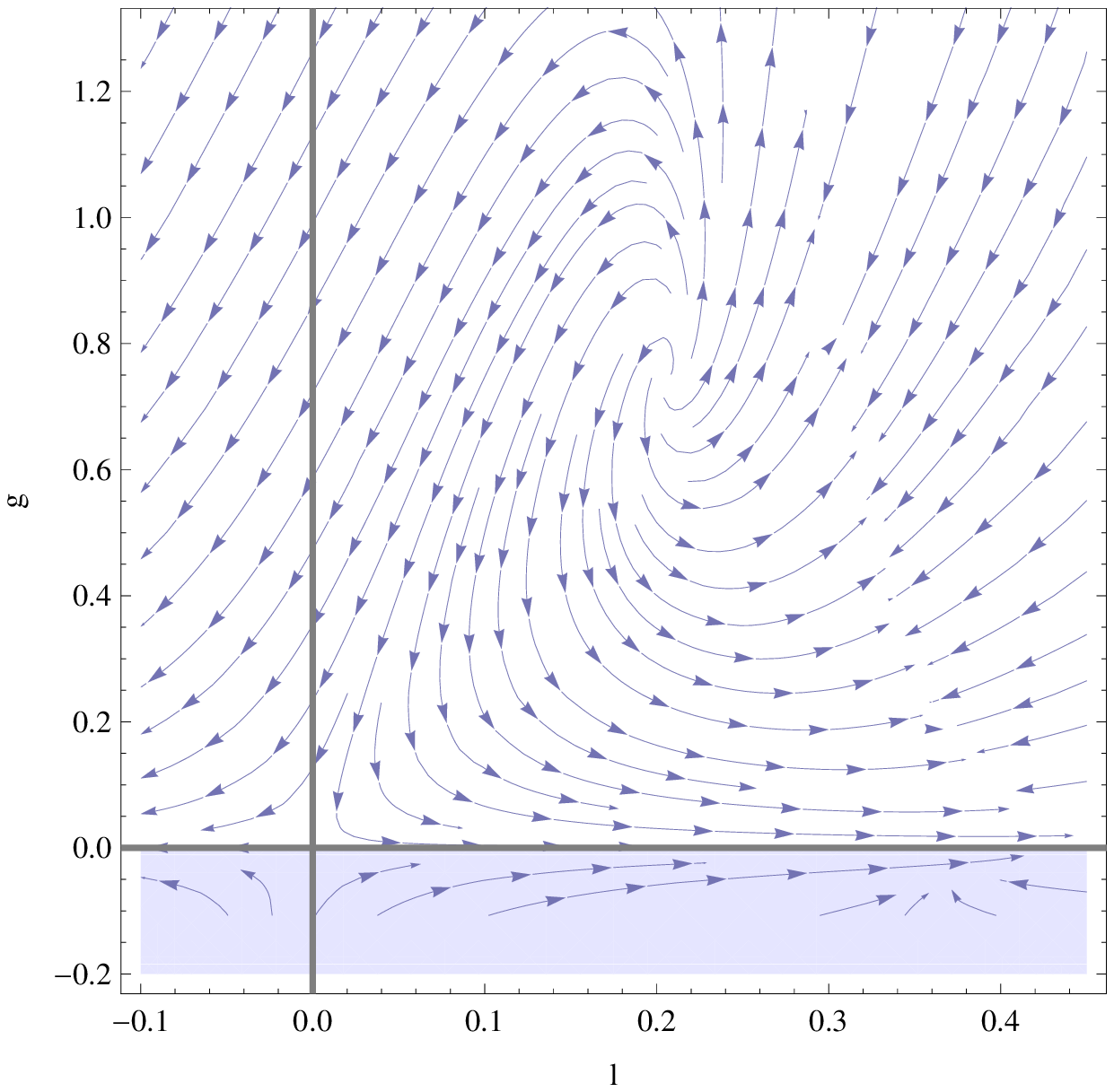}
 }
\caption{The  phase portraits for the dynamical and the level-(0) sector according to the beta-functions of [\Rmnum{2}]. As can be seen from the shading  the dynamical RG flow exhibits a transition from negative to positive anomalous dimensions at $\KkD_{\crit}>0$, while its level-(0) counterpart is negative everywhere in the physically relevant part of theory space.}  
 \label{fig:b1db1smBM}
\end{figure}

The property at stake is the $\KkD$-dependence of $\eta^{\dyn}$. 
It has the structure \eqref{eqn:res4D_005C} again, but with seemingly rather different  functions  in the numerator and denominator:
\begin{subequations}
\begin{align}
\kAD(\KkD)&=
  \frac{1}{\pi}\,\Big\{\tfrac{23}{3}  \, \ThrfA{2}{2}{-2\KkD}
 -24 \, \ThrfA{3}{3}{-2\KkD} 
 -\tfrac{2}{3} \ThrfA{2}{2}{0} + 8 \ThrfA{3}{3}{0} \Big\}
\label{eqn:res4D_006}
\\
\kBD(\KkD)&=
  -\frac{1}{2\pi}\,\Big\{\tfrac{23}{3}  \, \ThrfB{2}{2}{-2\KkD}
-24 \, \ThrfB{3}{3}{-2\KkD} 
 \, \Big\}
\label{eqn:res4D_007}
\end{align}
\end{subequations}
The level-(0) sector is governed by the following anomalous dimension $\eta^{(0)}$:
\begin{align}
 \eta^{(0)}(\tg^{\dyn},\KkD,\tg^{(0)})&=
 \frac{2}{\pi}\, \left[\frac{5}{6} \qA{1}{1}{-2\KkD}
- 3\, \qA{2}{2}{-2\KkD}-    \tfrac{2}{3}  \ThrfA{1}{1}{0}   - \,\ThrfA{2}{2}{0}\right]\tg^{(0)}
\label{eqn:res4D_003X}
\end{align}

\begin{figure}[ht!]
\centering
 \psfrag{c}[rb]{${\scriptstyle  \KkD_{\crit}}$}
  \psfrag{g}[c]{${\scriptstyle  \tg^{\dyn}}$}
   \psfrag{l}{${\scriptstyle \Kk^{\dyn}}$}
  \includegraphics[width=0.5\textwidth]{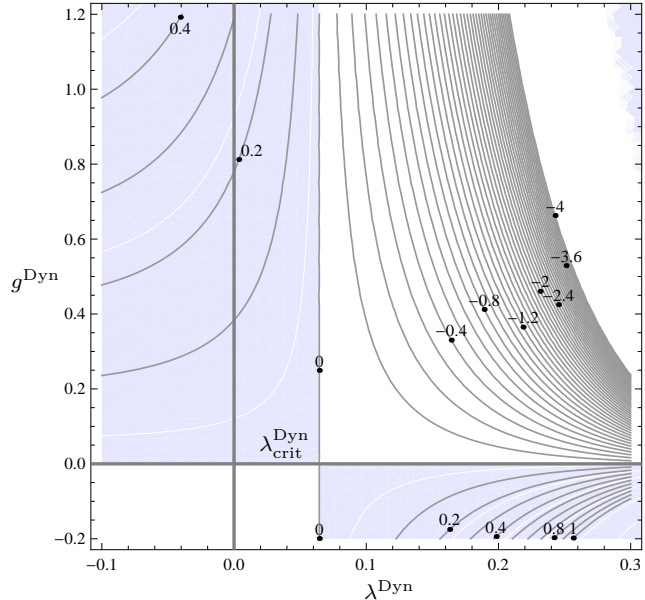}
\caption{The contour plot shows the lines of constant anomalous dimension $\eta^{\dyn}$ implied by the bi-metric calculation [\Rmnum{2}].
The shaded regions correspond to $\eta^{\dyn}>0$. Consistent with the calculation in [\Rmnum{1}],  $\eta^{\dyn}$ is found to change its sign on a straight line $\KkD=\KkD_{\crit}$ with $\KkD_{\crit}>0$. In the semi-classical regime, $\eta^{\dyn}$ is seen to be positive.}   \label{fig:etaContourBM}
\end{figure}

The resulting  phase-portraits for the dynamical and level-(0) sectors are depicted in Fig. \ref{fig:b1db1smBM}.
Only in case of the dynamical anomalous dimension $\eta^{\dyn}$ do the shaded areas, indicating regions of positive anomalous dimension, appear in the physically relevant part of the phase diagram. For the level-(0) sector we obtain a negative value of $\eta^{(0)}$ everywhere.
The contour plot over the $(\tg^{\dyn},\KkD)$ plane showing the lines of constant $\eta^{\dyn}$ is displayed in Fig. \ref{fig:etaContourBM}.

Comparing the diagrams in Figs. \ref{fig:b1db1smBM} and \ref{fig:etaContourBM} to their analogs of the calculation [\Rmnum{1}], in Figs. \ref{fig:b1db1smMRS2LD} and \ref{fig:etaContourMRS2}, {\it we find perfect agreement at the qualitative level between the two bi-metric approaches [\Rmnum{1}] and [\Rmnum{2}], respectively.
However, the results differ significantly from their single-metric counterparts in Fig. \ref{fig:etaSM}.}

As the semi-classical regime is of special importance let us expand  $B_1^{\dyn}(\KkD)$ in the vicinity of $\KkD=0$ again:
\begin{align}
B_1^{\dyn}(\KkD)&=\frac{1}{\pi}\Big[7\ThrfA{2}{2}{0}-16\ThrfA{3}{3}{0}\Big]-\frac{26}{3\pi}\KkD+\Order{(\KkD)^2}
\label{eqn:qeg020}
\end{align}
Eq. \eqref{eqn:qeg020} confirms the picture implied by the previous bi-metric calculation [\Rmnum{1}]:
a universal, negative slope (which in this case happens to coincide with the single-metric value, $-26\slash 3\pi$) along with a universally {\it positive} constant term, $B_1^{\dyn}(0)$.
Together they give rise to a region in which $\eta^{\dyn}>0$.

\begin{figure}[h!]
\psfrag{d}[bc][bc][1][90]{${\scriptstyle B_1^{\dyn}(\Kk;4)-B_1^{(0)}(\Kk;4)}$}
\psfrag{g}{${\scriptstyle  \tg^{(0)}}$}
 \psfrag{l}{${\scriptstyle \Kk^{(0)}}$}
\psfrag{b}[bc]{${\scriptstyle B_1^{\dyn}(0)}$}
\psfrag{a}[bc]{${\scriptstyle s}$}
\psfrag{t}[bc]{${\scriptstyle }$}
\psfrag{r}[bc]{${\scriptstyle }$}
\centering
\includegraphics[width=0.6\textwidth]{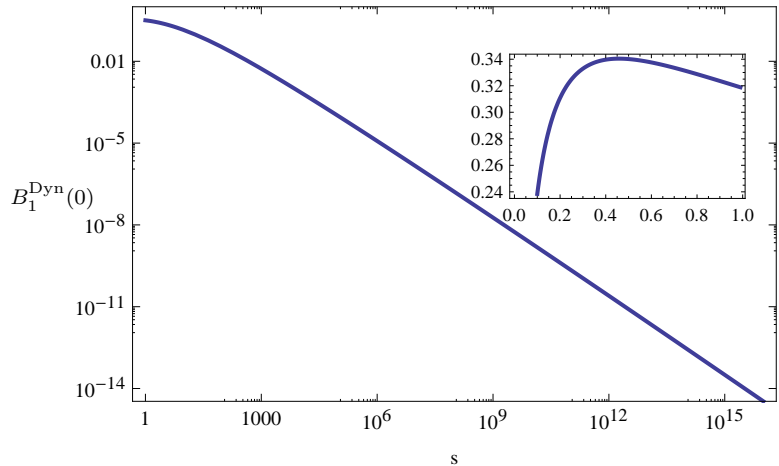}
\caption{The value of $B_1^{\dyn}(0)$ according to  the bi-metric calculation [\Rmnum{2}] is shown for different choices of the family parameter $s$ characterizing the exponential cutoff shape functions.}  \label{fig:b1db1sm}
\end{figure}
The cutoff-scheme dependence of $B_1^{\dyn}(0)$ was again checked  by evaluating $B_1^{\dyn}(0)$ for the optimized and the $s$-family of exponential shape-functions, for instance. 
The optimized cutoff yields $B_1^{\dyn}(0)=5\slash 6\pi$.
The linear approximation \eqref{eqn:qeg020} corresponds to the critical value  $\KkD_{\crit}\approx 0.096$ in this case; it coincides almost perfectly with the corresponding exact  value from the full non-linear equation: $\KkD_{\crit}\approx 0.064$.
For the exponential shape-functions the constant term in \eqref{eqn:qeg020} evaluates to 
$B_1^{\dyn}(0)=\frac{4-4 s+\ln(s)+3 s \ln(s)}{\pi  (s-1)^2}$ which is positive for all admissible values\footnote{It is known that for quantitative reliability the shape parameter $s\in(0,\infty)$ should not be chosen too small, $s\gtrsim0.5$, say \cite{frank1,oliver1,oliver2}.} of $s$, as shown in Fig. \ref{fig:b1db1sm}.

Thus, the second set of bi-metric results fully confirms all conclusions drawn in the previous subsection on the basis of the RG equations obtained in [\Rmnum{1}].

\subsection{Summary: significance of the cosmological constant}
We investigated the possibility of a positive anomalous dimension ($\eta^{\dyn}$ or $\eta^{\sm}$) in the semi-classical regime of three different truncations.
In the Introduction we discussed already that while at negative $\eta^{\cix}$ near the NGFP is the very hallmark of Asymptotic Safety, there is no general reason that would forbid $\eta^{\cix}$ to be positive in other parts of theory space, the semi-classical regime in particular.
While a transition to a positive $\eta^{\cix}$ was not observed in any single-metric truncation, we found that both bi-metric calculations which we analyzed do indeed show that $\eta^{\dyn}$ is actually positive on a large portion of theory space, namely the half plane $-\infty<\KkD<\KkD_{\crit}$.
Here $\KkD_{\crit}$ is a strictly positive critical cosmological constant, necessarily smaller than the NGFP coordinate $\KkD_*$.
\begin{figure}[h!]
\psfrag{d}[bc][bc][1][90]{${\scriptstyle B_1^{\dyn}(\Kk;4)-B_1^{(0)}(\Kk;4)}$}
\psfrag{g}{${\scriptstyle  \tg^{(0)}}$}
 \psfrag{l}{${\scriptstyle \Kk^{(0)}}$}
\psfrag{b}[bc]{${\scriptstyle B_1^{\cix}(\Kk)}$}
\psfrag{c}[tc]{${\scriptstyle {\rm[\Rmnum{1}]}}$}
\psfrag{d}[bc]{${\scriptstyle {\rm[\Rmnum{2}]}}$}
\psfrag{e}[bc]{${\scriptstyle\sm}$}
\psfrag{a}[bc]{${\scriptstyle \Kk}$}
\centering
\includegraphics[width=0.6\textwidth]{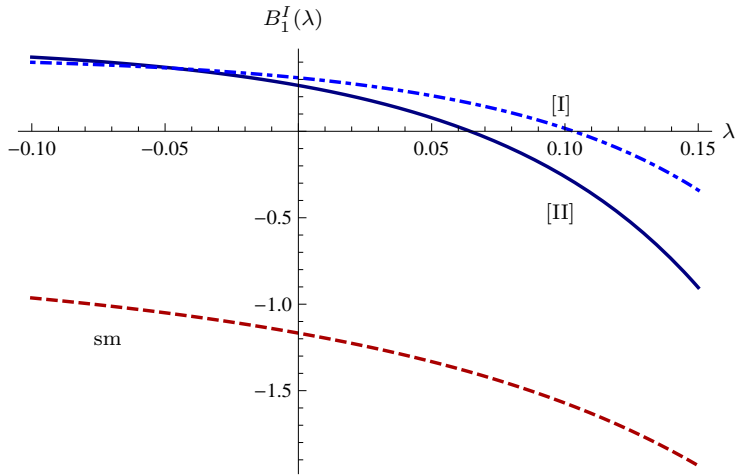}
\caption{The $\Kk$-dependence of $B_1^{\cix}(\Kk)$ for the two  bi-metric truncations [\Rmnum{1}] and [\Rmnum{2}], as well as the single-metric approximation ($\sm$). Notice that the latter has not only a negative slope but also a negative intercept $B_1^{\sm}(0)<0$, while both bi-metric functions are positive in the semi-classical regime of not too large dimensionless cosmological constant.}  \label{fig:b1_v1}
\end{figure}

The region in theory space with a negative $\eta^{\dyn}$, which is indispensable for  a non-Gaussian fixed point and the non-perturbative renormalizability of QEG, crucially owes its existence to  the {\it negative, universal slope} of $B_1^{\cix}(\Kk)$ at $\Kk=0$. 
It occurs in all three truncations, including the single-metric one, and indicates an {\it anti-screening} component in  the beta-function of $\tg^{\dyn}$.
In the `sm' case the intercept $B_1^{\sm}(0)$ is negative as well, and so $B_1^{\sm}(\Kk)$ is negative for all $\Kk$.
In both bi-metric truncations $B_1^{\dyn}(0)$ is positive, however, and this gives rise to a window $\KkD\in(-\infty,\KkD_{\crit})$ with a certain $\KkD_{\crit}>0$ in which $B_1^{\dyn}(\Kk)$ is positive.

In the semi-classical regime, the linear (in $\tg^{\dyn}$) relationship $\eta^{\dyn}\approx B_1^{\dyn}(\Kk)\tg^{\dyn}$ always turned out to be an excellent approximation. 
Hence, for a positive Newton constant (which we always assume) the anomalous dimension is positive in the window $\KkD\in (-\infty,\KkD_{\crit})$.
The precise value of $\KkD_{\crit}$ depends on the cutoff shape function; generically it is of the order $10^{-1}$ or $10^{-2}$, say.

The main message is summarized in Fig. \ref{fig:b1_v1} which depicts the exact (i.e., all-order) $\Kk$-dependence of $B_1^{\cix}$.
The single- and bi-metric functions all decrease with increasing $\Kk$.
But while the `$\sm$' function $B_1^{\sm}$ is negative everywhere, both of the dynamical bi-metric functions are non-negative in the vicinity of $\Kk=0$, implying a positive dynamical anomalous dimension there: $\eta^{\dyn}(\tg^{\dyn},\KkD)>0$  for all $\tg^{\dyn}>0$ and $-\infty<\KkD<\KkD_{\crit}$.

\subsection{From anti-screening to screening and back}

Recalling the definition $\eta^{\dyn}\equiv k\partial_k \ln G_k^{\dyn}$, it follows from the above that along every RG trajectory running on the half space $\KkD<\KkD_{\crit}$ the dynamical Newton constant $G_k^{\dyn}$ {\it increases} with increasing scale $k$.
Stated differently, the gravitational interaction shows a {\it screening} behavior there.
This is in stark contrast to its anti-screening character in the NGFP regime.

\begin{figure}[h!]
\psfrag{d}[bc][bc][1][90]{${\scriptstyle B_1^{\dyn}(\Kk;4)-B_1^{(0)}(\Kk;4)}$}
\psfrag{g}{${\scriptstyle  \tg^{\dyn}}$}
 \psfrag{l}[tx]{${\scriptstyle \Kk^{\dyn}}$}
\psfrag{A}[bc]{${\scriptstyle \tg_*^{\dyn}}$}
\psfrag{B}[bc]{${\scriptstyle \tg_T^{\dyn}}$}
\psfrag{C}[mc]{${\scriptstyle \KkD_T}$}
\psfrag{D}[mc]{${\scriptstyle \KkD_*}$}
\psfrag{k}[ml]{${\scriptstyle \KkD_{\crit}}$}
\psfrag{O}[mc]{${\scriptstyle O}$}
\psfrag{t}[bc]{${\scriptstyle T}$}
\psfrag{f}[l]{${\scriptstyle k_{\crit}^{\IR}}$}
\psfrag{u}[l]{${\scriptstyle k_{\crit}^{\UV}}$}
\centering
\includegraphics[width=0.6\textwidth]{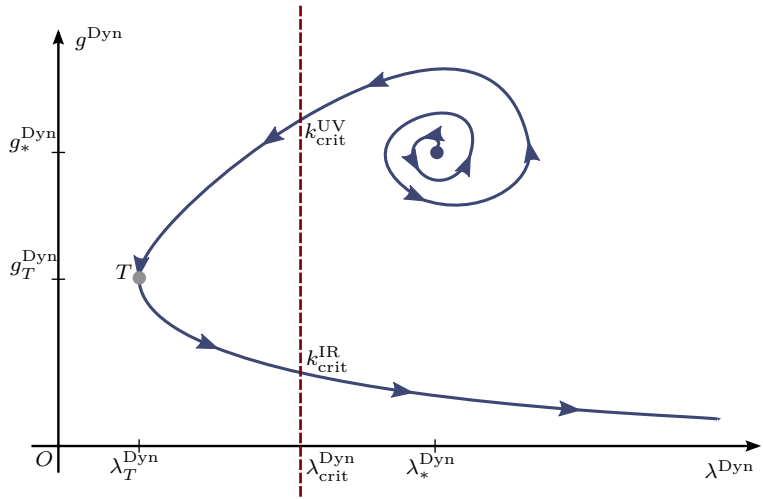}
\caption{Schematic behavior of a bi-metric type \Rmnum{3}a trajectory on the $(\tg^{\dyn},\KkD)$-projection of theory space. The dashed line separates the half spaces with $\eta^{\dyn}>0$ and $\eta^{\dyn}<0$, respectively. The part of the trajectory located above (below) the turning point $T$ is referred to as the trajectory's UV (IR) branch.}  \label{fig:scAntiScreen}
\end{figure}
For the example of a bi-metric trajectory which is of type \Rmnum{3}a in the `$\dyn$' projection \cite{daniel2} the situation is depicted schematically in Fig. \ref{fig:scAntiScreen}.
The trajectory $k\mapsto (\tg_k^{\dyn},\KkD_k)$ emanates from the NGFP at `$k=\infty$', then leaves the asymptotic scaling regime for $k\approx m_{\text{Pl}}$, but stays in the half-space with $\eta^{\dyn}>0$ as long as $k$ is larger than a certain critical scale $k_{\crit}^{\UV}$ at which the running cosmological constant $\KkD_{k}$ drops below $\KkD_{\crit}$.
As $k$ decreases further below $k^{\UV}_{\crit}$, the cosmological constant  continues to decrease until the turning point $T$ is reached, beyond which the (dimensionless!) $\KkD_k$ now increases for decreasing $k$.
Ultimately, it will re-enter the half-space with $\eta^{\dyn}<0$, namely at a second critical scale, $k_{\crit}^{\IR}$.
So, by definition,
\begin{align}
\KkD_k|_{k={k_{\crit}^{\IR}}}=\KkD_{\crit}=\KkD_k|_{k={k_{\crit}^{\UV}}} \quad \text{ with } k_{\crit}^{\IR}<k_{\crit}^{\UV}\,.
\end{align}

As it is already well-known for the type \Rmnum{3}a trajectories in the single-metric truncation \cite{frank1,h3,entropy}, the bi-metric trajectories of this type, too, can have a long classical regime where the (dimensionful!) Newton- and cosmological constant are approximately constant.
This requires tuning the turning point $T$ very close to the Gaussian fixed point, the origin $(0,0)$ in Fig. \ref{fig:scAntiScreen}.
The point $T$ is passed at $k=k_T$ with $k_{\crit}^{\IR}\ll k_T\ll k_{\crit}^{\UV}$ where the two critical scales are far apart then.

For example, the `RG trajectory realized in Nature', that is, the specific single-metric $(\tg^{\sm}_k,\Kk^{\sm}_k)$- or bi-metric $(\tg_k^{\dyn},\KkD_k)$-trajectory whose parameters are matched against the measured values of $G$ and $\Kkbar$ \cite{h3,entropy} is well-known to be highly fine-tuned, with turning point coordinates as tiny as $g_T\approx \Kk_T\approx 10^{-60}$.
Following the discussion in \cite{h3,entropy} it is easy to see that, for this trajectory, and for a $\KkD_{\crit}$ value of, say, $10^{-2}$, the UV critical scale is about $k_{\crit}^{\UV}\approx m_{\text{Pl}}\slash 10$, while the one in the IR is slightly above the present Hubble parameter, $k_{\crit}^{\IR}\approx 10 H_0$.
Newton's constant reaches its maximum at $k=k_{\crit}^{\UV}$; it is about $2\%$ larger there than at laboratory scales.

\section{Interpretation and Applications}
The `dynamical' anomalous dimension $\eta^{\dyn}$ governs the running of that particular version of Newton's constant which controls the strength of the gravitational self-interaction and the coupling of gravity to matter.
We found gravitational screening (rather than anti-screening, as predicted by the single-metric truncations) in the semi-classical regime, that is, $G_k^{\dyn}$ grows  with $k$ as long as $\KkD_k<\KkD_{\crit}$.
The strong renormalization effects associated with Asymptotic Safety, the formation of a fixed point, anti-screening, and  large negative values of $\eta^{\dyn}$, are confined to the half-space with $\KkD>\KkD_{\crit}$ instead.

In the following two subsections we discuss a number of possible implications of these findings.
In subsection \ref{sec:darkMatter} we interpret the sign change of $\eta^{\dyn}$ in terms of a dark matter description, and in subsection \ref{sec:appCosmo} we briefly comment on an application in cosmology.

\subsection{The dark matter interpretation} \label{sec:darkMatter}
\noindent{\bf(A) Physical significance of the dimensionless cosmological constant.}
For the interpretation of the above results it is helpful to recall that, upon going on-shell, the value of the {\it dimensionful} cosmological constant $\KkbarD\equiv k^2 \KkD_k$ determines the curvature of spacetime when it is explored with an experiment, or a `microscope' of resolving power\footnote{This estimate could be made more precise using the method of the `cutoff modes', see refs. \cite{jan1,jan2}.} $\ell\propto 1\slash k$. 
The radius of curvature of spacetime is of the order $r_c\propto \big(\KkbarD\big)^{-1\slash 2}$ then, and the {\it dimensionless} cosmological constant is approximately the (squared) ratio of the two distance scales involved:
\begin{align}
\KkD_{k}\approx \left(\frac{\ell}{r_c}\right)^2 
\label{eqn:inp_01}
\end{align}
Thus we see that {\it the sign-flip of $\eta^{\dyn}$ is controlled by the background curvature:} on self-consistent backgrounds \cite{daniel1} which are only weakly curved on the scale of the microscope, $\ell\ll r_{\rm c}$, we have $\KkD_{k}\ll 1$, therefore $\eta^{\dyn}>0$, and so we observe a screening behavior of the gravitational interaction.
Conversely, when the spacetime is strongly curved on the scale of the microscope (i.e. the scale set by the modes just being integrated out at this $k$) the ratio $\ell\slash r_{\rm c}$ approaches unity, implying $\KkD_k>\KkD_{\crit}$ and, as a result, strong anti-screening effects.

\noindent{\bf(B) Propagating gravitons in the semi-classical regime.}
The positive $\eta^{\dyn}$ in the semi-classical regime resolves the puzzle raised in the Introduction:
On a nearly flat background spacetime the dynamics of the $\flcb_{\mu\nu}$ fluctuations is such that the interactions get weaker at large distance, and the corresponding Green's function is short ranged.
The positive $\eta^{\dyn}$ causes no conflict with the existence of a K\"{a}llen-Lehmann representation with a positive spectral density, and the EAA may be seen as describing an effective field theory very similar to those on Minkowski space.
It describes weakly interacting gravitons and, in the classical limit, gravitational waves. 
In the opposite extreme when the curvature is large on the scale set by $k$ there is no description of the $\flcb_{\mu\nu}$-dynamics in terms of a Minkowski space-like effective field theory.
The propagator $\propto 1 \slash \left(-\bZ^2\right)^{1-\eta^{\dyn}\slash 2}$ is very different from the one on flat space then, both because of the background curvature and of the large negative $\eta^{\dyn}$ which renders it long ranged.
In this regime the $\flcb_{\mu\nu}$-dynamics is anti-screening and results in the formation of a non-trivial RG fixed point.

This general picture points in a similar direction as the mechanism of the `paramagnetic dominance' found in \cite{andi1} which likewise emphasizes the importance of the background curvature for Asymptotic Safety.

The positive sign of $\eta^{\dyn}$ near the Gaussian fixed point is furthermore consistent with the perturbative calculations on a flat background\footnote{Provided the latter are restricted to the vacuum-polarization diagrams, i.e. those related to $\eta^{\dyn}$.} performed by Bjerrum-Bohr, Donoghue, and Holstein \cite{donoghue-bjerrum}.

The screening behavior in the semi-classical regime is also consistent with the first analyses of the `lines of constant physics' \cite{CDT-const-phys,lines-ocph} found by numerical simulations within the CDT approach \cite{CDT-PhysRep}.

\noindent{\bf(C) Strong curvature regime: `physical', gravitating, and (non-)propagating $\bm{\flcb_{\mu\nu}}$ modes.}
An important issue about which we can only speculate at this point is the properties of the metric fluctuations in the regime where $\KkD \gtrapprox \KkD_{\crit}$.
There, the field $\flcb_{\mu\nu}$ still carries `physical', in the sense of `non-gauge' excitations which, however, admit no description as `particles' approximately governed by an effective field theory similar to those on Minkowski space.
This would not be surprising from an {\it on-shell} perspective as now the background is curved on a scale comparable to the physics considered. 
However, it is not completely trivial that the quantum fluctuations driving the RG flow\footnote{By contributing to the functional trace on the RHS of the FRGE.} reflect this transition, too, since those are far off-shell in general.

All we can say about the $\flcb_{\mu\nu}$ quantum field in this regime is that it is likely to carry `physical' excitations which, due to the non-linearity of the theory, interact gravitationally.
We do not know the precise propagation properties of those excitations, however. 
They might, or might not behave like a curved space version of the graviton, as propagating little ripples on a strongly curved background.

What comes to mind here is the analogy to transverse gluons in QCD, at the transition from the asymptotic freedom to the confinement regime.
In either regime they are `physical', i.e. `non-gauge' excitations, but only in the former regime they behave similar to  propagating particles, while they are confined in the latter.

Also the {\it unparticles} which were proposed by Georgi in a different context \cite{unparticles} are examples of such perfectly `physical' field excitations which  admit no particle interpretation, not even on flat space.

\noindent{\bf (D) The $\bm{\flcb_{\mu\nu}}$ propagator by RG improvement.}
The physics of the $\flcb_{\mu\nu}$ excitations in the strong curvature regime could be explored by computing their $n$-point functions $\delta^n \EAA_0[\flcb;\bg]\slash \delta \flcb^n|_{\flcb=0}$ from the standard effective action $\EAA_0=\lim_{k\rightarrow0}\EAA_k$ on a self-consistent, in general curved background $\bg\equiv\bg^{\sCon}$.
Particularly important is the inverse propagator ${\cal G}^{-1} \propto \delta^2\EAA_0 [\flcb;\bg^{\sCon}]\slash \delta \flcb^2|_{\flcb=0} $.
It describes the properties of both the `radiative' modes carried by $\flcb_{\mu\nu}$,  and the `Coulombic' modes. 
The latter determine in particular the response of the $\flcb_{\mu\nu}$ field to an externally prescribed (static) source ${\cal T}_{\mu\nu}$, the source-field relationship having the symbolic structure ${\cal G}^{-1}\flcb={\cal T}$.

The calculation of ${\cal G}$ is a very hard problem, not only because of the much more general truncation ansatz it requires, but also because we do not yet know any realistic candidate for a consistent background $\bg_k^{\sCon}$ in the domain of interest \cite{oliver+wetterich+mr,lasagne}.
Clearly a technically simple background like $\bg_{\mu\nu}=\delta_{\mu\nu}$ is excluded here since a flat background is far from consistent when $\KkD$ is large.

Despite these difficulties we can try to get a rough first impression of this domain if we restrict our attention to the $\flcb_{\mu\nu}$ propagator in a regime of covariant momenta in which $\eta^{\dyn}\equiv \eta$ is approximately $k$-independent.
Then, by a standard argument \cite{cwRGimpprop}, RG improvement of the 2-point function suggests that the inverse propagator in $\EAA_0$ equals ${\cal G}^{-1}\propto \left(-\bZ^2\right)^{1-\eta\slash2}$.
In general this is a complicated operator with a non-local integral kernel.
Let us consider the corresponding source-field relation, 
\begin{align}
L^{-\eta}\,\left(-\bZ^2\right)^{1-\eta\slash2}\,\phi=-4\pi G \rho
\label{eqn:inp_02}
\end{align}
with a now scale-independent Newton constant $G$, and a length parameter $L$ included for dimensional reasons.
Here we suppress the tensor structure and employ a notation reminiscent of the Newtonian limit which we shall take later on only; the following argument is fully relativistic still.

\noindent{\bf(E) Non-locality mimicks dark matter.}
For a generic real, i.e. non-integer value of $\eta$ the LHS of eq. \eqref{eqn:inp_02} involves a highly non-local operator acting on $\phi$.
In order to understand how the solutions of this equation differ from the classical ones, let us act with the operator $\big(-L^2 \bZ^2\big)^{\eta\slash2}$ on both sides of \eqref{eqn:inp_02}. 
Leaving domain issues aside this yields an equation similar to \eqref{eqn:inp_02}, but now with $\eta=0$ and a modified source instead:
\begin{subequations}
\begin{align}
\bZ^2 \phi& = 4\pi G \tilde{\rho} \label{eqn:inp_03A}\\
\tilde{\rho} &\equiv \big(-L^2 \bZ^2\big)^{\eta\slash2} \rho
\label{eqn:inp_03B}
\end{align}
\label{eqn:inp_03}
\end{subequations}
We see that the modifications caused by  a non-zero anomalous dimension can be shifted from the differential operator acting on the gravitational field to the source function.
In the Newtonian limit, for instance, eq. \eqref{eqn:inp_03A} has the interpretation of the classical Poisson equation for the graviational potential $\phi$ generated by the mass density $\tilde{\rho}$.
However, the density function $\tilde{\rho}$ does not coincide with  the mass distribution that has actually been externally prescribed, namely $\rho$.
The RG effects are  encoded in the way the `bare' mass distribution $\rho$ gets `dressed' by quantum effects which turn it into the `renormalized' $\tilde{\rho}$.

Being more explicit, the operator application in \eqref{eqn:inp_03B} amounts to the convolution of $\rho$ with a non-local integral kernel:\footnote{If needed, the non-integer power of $\bZ^2$ can be expressed by an appropriate integral representation. For a general discussion of fractional powers of the Laplacian and d'Alembertian and their Green's functions, see \cite{powers}.}
\begin{subequations}
\begin{align}
\tilde{\rho}(x)&= \int \md^d x^{\prime}\sqrt{\bg(x^{\prime})}\, K_{\eta}(x,x^{\prime})\,\rho(x^{\prime})
 \label{eqn:inp_04A}\\
K_{\eta}(x,x^{\prime})&\equiv \langle x|\big(-L^2 \bZ^2\big)^{\eta\slash 2}| x^{\prime} \rangle
\label{eqn:inp_04B}
\end{align}
\label{eqn:inp_04}
\end{subequations}
Note that the kernel $K_{\eta}$, and therefore $\tilde{\rho}$, still depend on the background $\bg_{\mu\nu}$.

While in general $x$ and $x^{\prime}$ are 4-dimensional coordinates they reduce to 3D space coordinates if we invoke the Newtonian limit where $\rho$, $\tilde{\rho}$, and $\phi$ are time independent.
In fact, to gain a rough, but qualitatively correct intuition for the `dressing' $\rho\mapsto \tilde{\rho}$, it suffices to consider the Newtonian limit, an approximately flat background in particular, but to maintain a non-zero value of $\eta$.
Then, with $\bg_{\mu\nu}=\eta_{\mu\nu}$, eq. \eqref{eqn:inp_03A} boils down to the time independent Poisson equation $\nabla^2 \phi= 4\pi G \tilde{\rho}$, and the kernel $K_{\eta}(\vec{x},\vec{x^{\prime}})\equiv K_{\eta}(|\vec{x}-\vec{x^{\prime}}|)$ is easily evaluated in the plane wave eigenbasis of the Laplacian on flat space, $\nabla^2$:
\begin{align}
K_{\eta}(r)&= \int \frac{\md^3 p}{(2\pi)^3}\, \left(L^2 \vec{p}^2\right)^{\eta\slash 2}\, e^{i\vec{p}\cdot \left(\vec{x}-\vec{x^{\prime}}\right)}\,, \qquad r\equiv |\vec{x}-\vec{x^{\prime}}|
\label{eqn:inp_05}
\end{align} 
Focusing on the simplest case, $\eta\in[-2,-1]$, this integral yields\footnote{For other values of $\eta$ we must introduce explicit distance or momentum cutoffs into the integral \eqref{eqn:inp_05} in order to take account of the fact that  the approximation ${\cal G}^{-1}\propto \left(\bZ^2\right)^{1-\eta\slash 2}$ with a constant value of $\eta$ is valid only in a restricted regime. Being interested in qualitative effects only we shall not do this here. One also has to be careful about delta-function singularities at the origin; in particular we have $K_0(\vec{x},\vec{x^{\prime}})=\delta(\vec{x}-\vec{x^{\prime}})$, as it should be. }, at $r\neq0$,
\begin{align}
K_{\eta}(r)&=-\left[4\pi \Gamma(-1-\eta) \cos(\tfrac{\pi}{2}\eta)\right]^{-1}\, \frac{L^{\eta}}{r^{3+\eta}} 
\label{eqn:inp_06}
\end{align} 
Now, even if the `bare' $\rho(\vec{x})$ is due to a point mass, for example, $\rho(\vec{x})=M\delta(\vec{x})$, the `renormalized' or `dressed' mass distribution amounts to an extended, smeared out cloud with a density profile $\tilde{\rho}(\vec{x})=M K_{\eta}(|\vec{x}|)$.
If \eqref{eqn:inp_06} applies, $\tilde{\rho}$ has support also away from $\vec{x}=0$, falling off according to the power law
\begin{align}
\tilde{\rho}(r)\propto 1 \slash r^{3+\eta}\qquad \qquad (r>0)
\label{eqn:inp_07}
\end{align}
If $\eta$ is negative, the $\tilde{\rho}$ distribution is the more extended the larger is $|\eta|$.

While strictly speaking \eqref{eqn:inp_07} is valid only for $\eta\in [-2,-1]$, it highlights the main impact a negative $\eta$ has on gravity, also beyond the Newtonian limit:
If one sticks to the classical form of the field equation (here: Poisson's equation) the gravitational field  is sourced not only by the energy momentum tensor of the true matter (here: $\rho$) but in addition by a fictitious energy-momentum-, and in particular mass-distribution ($\tilde{\rho}$) which is obtained by a non-local integral transformation applied to the true, or `bare', source.

In the simplest case the integral transformation is linear and assumes the form \eqref{eqn:inp_04A}.
Where it applies, the `fictitious' matter traces the `genuine' one,
the latter sources the former.
Hence it seems indeed appropriate to regard the transition from $\rho$ to $\tilde{\rho}$ as due to the `dressing' of the bare source by quantum effects, similar to the dressing of electrons in QED by  clouds of virtual particles surrounding them.
It is quite clear then, in particular in a massless theory, that the dressing of point sources results in spatially extended, non-local structures.

\noindent {\bf(F) Modified gravity in astrophysics: a digression.}
Applying this discussion to the realm of astrophysics, to galaxies or clusters of galaxies, one is tempted to interpret the fictitious matter contained in $\tilde{\rho}$, over and above the true one, as the long sought-for dark matter, and to identify $\rho$ with the actually observed `luminous' matter.

To avoid any misunderstanding we emphasize that the presently available RG flows do not (yet?) reliably predict large negative anomalous dimension ($\eta\approx -1$, say) on astrophysical scales\footnote{See, however, ref. \cite{h3}.}.
All we can say for the time being is that the mathematical structure of the field equations we encounter here is {\it potentially} relevant to the astrophysical dark matter problem, but clearly much more work will be needed to settle the issue.

The much more direct reason why the mechanism of non-local gravity mimicking dark matter is relevant to Asymptotic Safety is that {\it on a type \Rmnum{3}a trajectory large negative $\eta$'s  occur in two regimes:}
not only at astrophysical or cosmological scales, $k\lesssim k_{\crit}^{\IR}$, but also near the Planck regime, $k\gtrsim k_{\crit}^{\UV}$.

As it is shown schematically in Fig. \ref{fig:scAntiScreen}, the trajectories of type \Rmnum{3}a, like the one that could perhaps apply to the real Universe, have two sections with a sufficiently large $\KkD$ to make $\eta^{\dyn}$ negative, one on  the UV-, the other on the IR-branch.
The main difference between the branches is their typical value of $\tg^{\dyn}$: it is much smaller on the IR-branch than on the UV-branch.
As a result, on the IR-branch $|\eta^{\dyn}|=|B_1^{\dyn}(\KkD)\,\tg^{\dyn}|$ assumes values of order unity, say, only when $\KkD$ is increased much further beyond $\KkD_{\crit}$ than this would be necessary on the UV-branch.
This distinction is best seen in the contour plots (`iso-$\eta$-lines') of Figs. \ref{fig:etaContourMRS2} and \ref{fig:etaContourBM}.
Since the Einstein-Hilbert truncation becomes unreliable near $\KkD=1\slash 2$, it can deal with the large negative $\eta$'s on the UV-branch only.

After the above precautionary remark it is nevertheless interesting to note that on the astrophysical side an integral transform like \eqref{eqn:inp_04A}, connecting luminous to dark matter in real galaxies, has indeed been proposed long ago on a purely phenomenological basis:
It is at the heart of the Tohline-Kuhn modified-gravity approach \cite{tohline, kuhn, bekenstein-review}.
Recently this approach has attracted  attention also because it  was found to emerge naturally from a certain classical, fully relativistic, and  {\it non-local} extension of General Relativity \cite{hehl-mashhoon}.

Above we saw that  quantum gravity effects can modify Einstein's  equations in precisely the Tohline-Kuhn style.
The similarity between the two theories becomes most explicit for $\eta=-1$ which leads to the integral kernel
\begin{align}
K_{-1}(\vec{x},\vec{x^{\prime}})&=\frac{1}{2\pi^2\,L}\, \frac{1}{|\vec{x}-\vec{x^{\prime}}|^2}
\label{eqn:inp_08}
\end{align}
This is exactly the one which appears also in the Tohline-Kuhn framework.

Using this kernel in eq. \eqref{eqn:inp_04A}, a point mass with $\rho(\vec{x})=M\delta(\vec{x})$ is seen to surround itself with a spherical `dark matter halo' whose radial density profile is given by $\tilde{\rho}(r)=M\slash (2\pi^2 L\, r^2)$.
By virtue of $\nabla^2 \phi=4\pi G\tilde{\rho}$, this dark matter distribution generates the logarithmic potential $\phi(r)=(2GM\slash \pi L)\ln(r)$. 
In the Newtonian limit, it is well known to yield a perfectly flat rotation curve, that is, a test particle on a circular orbit has a velocity which is independent of its radius\footnote{From the idealized case of a point particle where $v^2\propto M$ it might appear that this approach has difficulties reproducing the Tully-Fisher law for spiral galaxies \cite{tully} which favors $v^4\propto M$. However, the detailed studies on the basis of fits to realistic galaxy data reported in \cite{mashhoon1,mashhoon2} seem  not to encounter such difficulties.}, $v^2=2GM\slash \pi L$.

\noindent {\bf(G) Non-local constitutive relations as a QEG vacuum effect.}
Recently the Tohline-Kuhn framework  turned out to describe the Newtonian limit of a fully relativistic generalization of General Relativity which allows the incorporation of non-locality at a phenomenological, purely classical level \cite{hehl-mashhoon,hehl-tele}.
This theory, proposed by Hehl and Mashhoon, relies on the observation that the teleparallel equivalent of General Relativity, a special gauge theory of the translation group, is amenable to generalization through the introduction of a non-trivial `constitutive relation' similar to the constitutive relations between $(\vec{E},\vec{B})$ and $(\vec{D},\vec{H})$ in electrodynamics.

Because of memory effects, such relations are non-local typically.
They make their appearance both in the classical electrodynamics of matter, and in vacuum Quantum Electrodynamics where loop effects are well known to give rise to a complicated relationship between $\vec{E}$ and $\vec{D}$, say, which is both non-linear and non-local \cite{dr1}.
As for quantum gravity, it  was pointed out  \cite{andi1, andi-MG} that QEG, like QED, has a non-trivial vacuum structure with a {\it non-linear} relationship between the gravitational analogs of the $\vec{E}$ and $\vec{D}$ fields.
From this perspective it is quite natural  that the   source-field relation of quantum gravity, in a regime with large negative $\eta$, turns out not only non-linear, but also {\it non-local}.

In this sense, a phenomenological theory like the one in \cite{hehl-mashhoon}, as far as its general structure is concerned,  may well be regarded as an effective field theory description of the QEG vacuum in the large-$\eta$ regime.

In fact, in QEG and the theory of ref. \cite{hehl-mashhoon} the size of the new effects is determined by essentially the same control parameter.
In \cite{hehl-mashhoon} the  degree of non-locality is governed by the ratio $\varrho\equiv L_{\text{acc}}\slash L_{\text{phen}}$, where $L_{\text{phen}}$ denotes the length scale of the phenomenon under consideration, and $L_{\text{acc}}$ is the acceleration length of the observer.
Interestingly, $\varrho^{-2}\equiv \left(L_{\text{phen}}\slash L_{\text{acc}}\right)^2$ is basically the same as the dimensionless cosmological constant $\Kk_k=\Kkbar_k\slash k^2$ which controls the size of $\eta$ and the non-local effects in QEG.
There, $\ell\approx k^{-1}$  characterizes the length scale of the physical process under consideration and so it takes the place of $L_{\text{phen}}$, while the radius of curvature, $r_{\rm c}$, may be identified with $L_{\text{acc}}$\footnote{In cosmology, for instance, one has indeed $L_{\text{acc}}\sim H^{-1}\sim\Kkbar^{-1\slash2}\sim r_{\rm c}$.}.

\noindent {\bf(H) Planck scale non-locality as `dark matter'.}
At this point of the discussion we switch back from the IR to the UV regime.
As we emphasized already the beta-functions considered in the present paper, where they are reliable, yield only tiny values for $\eta$ on astrophysical scales. 
So here we focus on the dark matter interpretation which  applies   to the {\it UV branch} of the `RG trajectory realized in Nature', see Fig \ref{fig:scAntiScreen}.
Of course, the UV-branch exists not only for the trajectories of type \Rmnum{3}a but for all asymptotically safe ones.
Along any of them, for $k$ near the Planck scale, but still above $k^{\UV}_{\crit}$, the anomalous dimension is large and  negative since the trajectory just left the NGFP regime where $\eta^{\dyn}\approx \eta^{\dyn}_*=-2$.

It is thus plausible to re-apply the above discussion of astrophysical dark matter which is mimicked by non-locality in the ultraviolet.
The situation would then be as follows.
When we approach the UV regime, above a certain scale $k^{\UV}_{\crit}$ located about one or two orders of magnitude below the Planck scale, non-local effects start  becoming essential.
Now, the regime in question, $k_{\crit}^{\UV}\lesssim k \lesssim m_{\text{Pl}}$,  is  exactly the one  for which we concluded already that the $\flcb_{\mu\nu}$ excitations cannot be described there by an effective field theory of the conventional local form; in particular their propagation properties are not easily established, and we conjectured that there are indeed no propagating gravitons above $k_{\crit}^{\UV}$.

Assuming this picture is correct it suggests  the interpretation of the physical, but non-propagating $\flcb_{\mu\nu}$ modes as a type of Planckian dark matter that admits an effective description in terms of a (fully relativistic!) Hehl-Mashoon-type theory \cite{hehl-mashhoon}.
In this scenario the modes of the metric fluctuations with covariant momenta above $k_{\crit}^{\UV}$ do not propagate, but are still physical (in the sense of `non-gauge').
They interact gravitationally with matter and among themselves, they can condense to form spatially extended structures, and they dress ordinary localized energy-momentum distributions by `dark matter halos' which are approximately described by  a Tohline-Kuhn-type integral transform.

This is the antagonism between gravitons and dark matter the title of this paper is alluding to:
The semi-classical modes of the fluctuation field have a particle interpretation, describe massless gravitons or essentially classical gravitational waves, while those with larger momenta are equally physical, gravitate, but do not propagate presumably.

To visualize this situation it helps to recall the example of the transverse gluon modes in QCD:
Those with momenta well above the confinement scale propagate approximately particle-like, the others are confined, and they form the homogeneous gluon condensate characteristic of the QCD vacuum state.

\subsection{Primordial density perturbations from the NGFP regime}\label{sec:appCosmo}
The conjectured absence of propagating gravitons in a certain range of momenta can  also be relevant to cosmology presumably, for example in the context of the cosmological microwave background radiation (CMBR).
In refs. \cite{cosmo1,entropy} an Asymptotic Safety-based alternative to the standard inflationary paradigm has been proposed in which the source of the primordial density perturbations, responsible for later structure formation, are the quantum fluctuations of geometry itself which occur during the Planck epoch.\footnote{For a different approach to asymptotically safe inflation see \cite{weinberg-AS-infl} and \cite{alfio-As-infl}.}
Within QEG the fluctuations in this regime are governed by the NGFP, and so they could provide a perfect window to the very physics of Asymptotic Safety. 

It has been argued that when the Universe was in the Planck-, or NGFP-regime the scale-free form of the  $\flcb_{\mu\nu}$-propagator $\propto 1\slash \Z^4$ gave rise to a kind of cosmic `critical phenomenon' which displays metric fluctuations on all length scales \cite{cosmo1, entropy, Naxos-book}.
The scale-free nature of all physics at the fixed point renders the fluctuation spectrum scale-free automatically.
Towards the end of the Planck era, the RG trajectory leaves the asymptotic scaling regime of the NGFP, the fluctuations `freeze out', and thus prepare the initial state for the subsequent classical evolution.
They lead to a Harrison-Zeldovich like CMBR spectrum with a spectral index of $n_s=1$ plus small corrections \cite{cosmo1, entropy, Naxos-book}.

Here the absence of propagating gravitational waves at high scales could come into play as follows.
At the end of the Planck epoch the geometry fluctuations get imprinted on the (by then essentially classical) spacetime metric and the matter fields.
The imprints then evolve classically, and ultimately, at decoupling, get encoded in the CMBR.
Now, a priori the frozen-in geometry perturbations present at the end of the Planck era ($k\approx m_{\text{pl}}$) would affect  the scalar and the radiative (`tensor') parts of the metric alike.
If, however, there do not yet exist physical radiative excitations  at this scale, or they are suppressed, then one has a natural reason to expect that in real Nature {\it the CMBR tensor-to-scalar-ratio should be  smaller than unity}.
The power in the tensor modes is suppressed relative to the scalar ones since by the time the Universe leaves the fixed point regime gravitational waves cannot propagate yet, the relevant scales being in the range $m_{\text{Pl}}>k>k^{\UV}_{\crit}$.

For the time being this is a somewhat speculative argument of course.
However, it is reassuring to see that it points in exactly the same direction as the  observational data on the tensor-to-scalar-ratio \cite{planck,bicep}.
\section{Summary}
Since the early investigations of the Einstein-Hilbert truncation it was clear that a subset of its RG trajectories contain a long classical regime at low scales in which $G_k$ and $\Kkbar_k$ are constant to a very good approximation; from these single-metric calculations it appeared, however, that in the adjacent semi-classical regime at slightly larger scales the Newton constant  decreases immediately, thus rendering the anomalous dimension $\eta\equiv k\partial_k \ln G_k$ negative.
Even though at the endpoint of the separatrix, for example, we have $\Kkbar=0$ and so the effective field equations admit Minkowski space as a solution, the quantized metric fluctuations on this background, the gravitons, would have unexpected properties, being more similar to gluons than to photons. 
However, in the present paper we provided evidence from two independent bi-metric analyses which indicate that this is actually not the case.
Between the strictly classical ($\eta=0$) and the fixed point regime ($\eta<0$) there exists an intermediate interval of scales with a positive anomalous dimension.
Those RG  trajectories which have a  positive cosmological constant in the classical domain possess two regimes displaying a negative anomalous dimension, one at Planckian, and the other on cosmological scales.
At least in the former the existence of propagating gravitons seems questionable, and we proposed a natural interpretation of the pertinent physical, non-propagating, but gravitating $\flcb_{\mu\nu}$ excitations as a form of Planckian `dark matter'.


\begin{thebibliography}{999}
\bibitem{donoghueEQFT}
 J.~F.~Donoghue, Phys. Rev. Lett. 72 (1994) 2996; Phys. Rev. D 50 (1994) 3874.

\bibitem{wein}
S.~Weinberg in {\it General Relativity, an Einstein
 Centenary Survey}, S.W.~Hawking and W.~Israel (Eds.), Cambridge
 University Press (1979).


\bibitem{mr}
M.~Reuter, Phys. Rev. D 57 (1998) 971, and
 {hep-th\slash9605030}.


\bibitem{frank1}
M.~Reuter and F.~Saueressig, Phys. Rev. D 65 (2002)
 065016\\ and {hep-th\slash 0110054}.


\bibitem{frank2}
M.~Reuter and F.~Saueressig,
Phys. Rev. D 66 (2002) 125001, hep-th\slash 0206145.


\bibitem{frank+friends}
E.~Manrique, S.~Rechenberger and F.~Saueressig, Phys. Rev. Lett. 106 (2011) 251302, \mbox{arXiv:1102.5012}.


\bibitem{frank-sig}
S.~Rechenberger and F.~Saueressig, JHEP 1312 (2013) 017.


\bibitem{frankfrac}
M.~Reuter and F.~Saueressig, JHEP 12 (2011) 012, \mbox{arXiv:1110.5224}.


\bibitem{frank-fR}
M.~Demmel, F.~Saueressig and O.~Zanusso,
JHEP 1211 (2012) 131; JHEP 1406 (2014) 026.



\bibitem{frank-ghost}
K.~Groh and F.~Saueressig,
J.\ Phys.\ A {\bf 43} (2010) 365403, \mbox{arXiv:1001.5032}. 


\bibitem{astrid-ghost}
A.~Eichhorn, H.~Gies and M.~M.~Scherer, Phys.\ Rev.\ D {80}, 104003 (2009);\\
A.~Eichhorn and H.~Gies, Phys.\ Rev.\ D {81}, 104010 (2010).


\bibitem{oliver1}
O.~Lauscher and M.~Reuter, Phys. Rev. D 65 (2002)
 025013, and \\{hep-th\slash 0108040}.


\bibitem{oliver2}
O.~Lauscher and M.~Reuter, Phys. Rev. D 66 (2002)
 025026, and \\{hep-th\slash 0205062}.


\bibitem{oliver3}
O.~Lauscher and M.~Reuter,
Class.\ Quant.\ Grav.\ 19 (2002) 483, \mbox{hep-th/0110021}.
%


\bibitem{oliver4}
O.~Lauscher and M.~Reuter,
Int.\ J.\ Mod.\ Phys.\ A17 (2002) 993, \mbox{hep-th/0112089}.

\bibitem{oliverfrac}
O.~Lauscher and M.~Reuter, JHEP\ 10 (2005) 050, \mbox{hep-th/0508202}.


\bibitem{perper}
R.~Percacci and D.~Perini,
Phys.\ Rev.\ D 67 (2003) 081503, \mbox{hep-th/0207033};
Phys.\ Rev.\ D 68 (2003) 044018, \mbox{hep-th/0304222}.



\bibitem{prop}
A.~Bonanno and M.~Reuter, JHEP\ 02 (2005) 035, \mbox{hep-th/0410191}.


\bibitem{elisa1}
E.~Manrique and M.~Reuter, Phys. Rev. D 79 (2009)
 025008\\ and {arXiv:0811.3888}.


\bibitem{vacca}
G.~P.~Vacca and O.~Zanusso, Phys. Rev. Lett. 105 (2010) 231601;\\O.~Zanusso, L.~Zambelli, G.~P.~Vacca and R.~Percacci, Phys. Lett. B 689 (2010) 90.


\bibitem{max-pert}
M.~Niedermaier, Phys. Rev. Lett. 103 (2009) 101303;\\
Nucl. Phys. B 833 (2010) 226.


\bibitem{creh1}
M.~Reuter and H.~Weyer, Phys.\ Rev.\ D 79 (2009) 105005, \mbox{arXiv:0801.3287};\\ Gen.\ Rel.\ Grav 41
(2009) 983, \mbox{arXiv:0903.2971}.


\bibitem{percacci}
R.~Percacci, \mbox{arXiv:1110.6389};\\
 P.~Dona, A.~Eichhorn, R.~Percacci, \mbox{arXiv:1311.2898}.


\bibitem{percadou}
D.~Dou and R.~Percacci,
Class.\ Quant.\ Grav.\ 15 (1998) 3449, \mbox{hep-th/9707239}.


\bibitem{percacci-pagani}

C.~Pagani and R.~Percacci, \mbox{arXiv:1312.7767}.

\bibitem{codello}
A.~Codello, R.~Percacci and C.~Rahmede, Ann. Phys.
 324 (2009) 414.


\bibitem{JE1}
J.~E.~Daum and M.~Reuter, Adv.\ Sci.\ Lett.\ 2 (2009) 255, \mbox{arXiv:0806.3907}.
%

\bibitem{JEUM}
J.-E.~Daum, U.~Harst and M.~Reuter, JHEP 01 (2010) 084, \mbox{arXiv:0910.4938}.


\bibitem{JEe-omega}
 J.~E.~Daum and M.~Reuter, Phys. Lett. B 710 (2012) 215 and \mbox{arXiv:1012.4280};
Annals Phys. 334 (2013) 351, \mbox{arXiv:1301.5135};
PoS (CNCFG 2010) 003 and \mbox{arXiv:1111.1000}.


\bibitem{QEG+QED}
U.~Harst and M.~Reuter, JHEP 05 (2011) 119 and
 {arXiv:1101.6007}.


\bibitem{andi1}
A.~Nink and M.~Reuter, JHEP\ 1301 (2013) 062, \mbox{arXiv:1208.0031}.

\bibitem{daniel1}
D.~Becker and M.~Reuter,
JHEP 07 (2012) 172 and \mbox{arXiv:1205.3583}.


\bibitem{daniel-MG}
D.~Becker and M.~Reuter, 
\mbox{arXiv:0802.2527}.


\bibitem{NJP}
For a recent review on QEG and Asymptotic Safety and a comprehensive list of references see M.~Reuter and F.~Saueressig, New J.Phys. 14 (2012) 055022 and \mbox{arXiv:1202.2274};\\Further details can be found in the {\it New Journal of Physics} special issue on QEG, http://iopscience.iop.org/1367-2630/focus/Focus on Quantum Einstein Gravity.


\bibitem{livrev}
 M.~Niedermaier and M.~Reuter, Living Reviews in Relativity 9 (2006) 5; \\
 M.~Reuter and F.~Saueressig, in {\it Geometric and Topological Methods for Quantum Field Theory}, H.~Ocampo, S.~Paycha and
 A.~Vargas (Eds.), Cambridge Univ.\ Press, Cambridge, 2010, \mbox{arXiv:0708.1317};\\
 R.~Percacci, in \textit{Approaches to Quantum Gravity: Towards a New Understanding of Space, Time and Matter}, D. Oriti (Ed.), Cambridge University Press, Cambridge, 2009,
 \mbox{arXiv:0709.3851}.


\bibitem{h3}
M.~Reuter and H.~Weyer,
JCAP\ 12 (2004) 001,
\mbox{hep-th/0410119}.


\bibitem{entropy}
A.\ Bonanno and M.\ Reuter, JCAP {08} (2007) 024, \mbox{arXiv:0706.0174}.


\bibitem{Ka-Leh}
 G.~K\"{a}ll\'{e}n, Helv. Phys. Acta 25 (1952) 417;\\ H.~Lehmann, Il Nuovo Cimento 11 (1954) 342.

\bibitem{nakanishi-ojima}
 N.~Nakanishi and I.~Ojima, {\it Covariant operator formalism of gauge theories and quantum gravity}, World Scientific, Singapore, 1990.

\bibitem{wett-mr}
M.~Reuter and C.~Wetterich, Nucl. Phys. B 391 (1993) 147,\\ Nucl. Phys. B 408 (1993) 91; C.~Wetterich, Phys. Lett. B 301 (1993) 90; \\M.~Reuter and C.~Wetterich, Nucl. Phys. B 417 (1994) 181,\\ Nucl. Phys. B 427 (1994) 291.


\bibitem{realQCD}
 A.~Maas, Phys. Rep. 524 (2013) 203;\\
L.~Fister, J.M.~Pawlowski, Phys. Rev. D 88 (2013) 045010;\\
S.~Strauss, C.S.~Fischer and C.~Kellermann, Phys. Rev. Lett. 109 (2012) 252001;\\
V.~Mader, M.~Schaden, D.~Zwanziger and R.~Alkofer, Eur. Phys. J. C 74 (2014) 2881.


\bibitem{west}

G.B.~West, Phys. Lett. B 115 (1982) 468.


\bibitem{elisa2}
E.~Manrique and M.~Reuter, Annals Phys.\ 325 (2010) 785, \mbox{arXiv:0907.2617}.


\bibitem{MRS1}
E.~Manrique, M.~Reuter and F.~Saueressig, Annals Phys. 326 (2011) 440, \mbox{arXiv:1003,5129}.


\bibitem{MRS2}
E.~Manrique, M.~Reuter and F.~Saueressig, Annals Phys. 326 (2011) 463, \mbox{arXiv:1006.0099}.


\bibitem{daniel2}
D.~Becker and M.~Reuter,
Annals Phys. 350 (2014) 225-301, \mbox{arXiv:1404.4537}.


\bibitem{Mottola}
 E.~Mottola, J. Math. Phys. 36 (1995) 2470-2511.


\bibitem{anharmRoberto}
  A.~Codello and R.~Percacci, Phys. Rev. Lett. 97 (2006) 221301;
A.~Codello, R.~Percacci and C.~Rahmede, Int. J. Mod. Phys. A 23 (2008).


\bibitem{frankmach}
P.~Machado and F.~Saueressig,
Phys.\ Rev.\ D 77 (2008) 124045, {\tt \mbox{arXiv:0712.0445}}.


\bibitem{donkin-paw}
 I.~Donkin and J.~M.~Pawlowski, \mbox{arXiv:1203.4207}

\bibitem{paw-rodigast}
N.~Christiansen, B.~Knorr, J.~M.~Pawlowski, A.~Rodigast, \mbox{arXiv:1403.1232}.


\bibitem{codello-closure}
A.~Codello, G.~D'Odorico, and C.~Pagani, \mbox{arXiv:1304.4777}.


\bibitem{matterPerc}
 P.~Dona, A.~Eichhorn, R.~Percacci, \mbox{arXiv:1311.2898}.


\bibitem{morris-dietz}
I. Hamzaan Bridle, J.~A.~Dietz, T.~R.~Morris, \mbox{arXiv:1312.2846}.

\bibitem{litimPRL}

D.~Litim, Phys.\ Rev.\ Lett.\ 92 (2004) 201301.


\bibitem{Saueressig:2011vn}
F.~Saueressig, K.~Groh, S.~Rechenberger and O.~Zanusso,\\
PoS EPS-HEP2011 (2011) 124 and \mbox{arXiv:1111.1743}.


\bibitem{BMS}
D.~Benedetti, P.~Machado and F.~Saueressig,
Mod.\ Phys.\ Lett.\ A {24} (2009) 2233, {\mbox{arXiv:0901.2984}}.


\bibitem{BMS2}
D.~Benedetti, P.~Machado and F.~Saueressig,
Nucl.\ Phys.\ B 824 (2010) 168, {\mbox{arXiv:0902.4630}}; {\mbox{arXiv:0909.3265}}.


\bibitem{creh2}
M.~Reuter and H.~Weyer,
Phys. Rev. D 80 (2009) 025001\\ and \mbox{arXiv:0804.1475}.


\bibitem{creh3}
P.~F.~Machado and R.~Percacci,
Phys.\ Rev.\ D 80 (2009) 024020, \mbox{arXiv:0904.2510}. 


\bibitem{stefan-frankfrac}
S.~Rechenberger and F.~Saueressig, Phys.\ Rev. D 86 024018 (2012), \mbox{arXiv:1206.0657}. 


\bibitem{morris-dietz-fR}
 J.~A.~Dietz and T.~R.~Morris, JHEP 1301 (2013) 108.


\bibitem{Benedetti-relevant}
 D.~Benedetti, Europhys. Lett. 102 (2013) 20007.

\bibitem{ohtaPerc}
 N.~Ohta and R.~Percacci, Class. Quant. Grav. 31 (2014) 015024.

\bibitem{souma}
W.~Souma,
Prog.\ Theor.\ Phys.\ 102 (1999) 181, \mbox{hep-th/9907027}.


\bibitem{jan1}
M.~Reuter and J.~Schwindt, JHEP\ 01 (2006) 070, \mbox{hep-th/0511021}.
%


\bibitem{jan2}
M.~Reuter and J.~Schwindt, JHEP\ 01 (2007) 049, \mbox{hep-th/0611294}.
%


\bibitem{donoghue-bjerrum}
 N.~E.~J.~Bjerrum-Bohr,
J.~F.~Donoghue, B.~R.~Holstein,\\ Phys. Rev. D67 (2003) 084033.

\bibitem{CDT-const-phys}
J.~Ambj\o rn, private communication.


\bibitem{lines-ocph}
 J.~Ambj\o rn, A.~Goerlich, J.~Jurkiewicz, A.~Kreienbuehl and\\ R.~Loll, \mbox{arXiv:1405.4585}.

\bibitem{CDT-PhysRep}
J.~Ambj\o rn, A.~Goerlich, J.~Jurkiewicz and R.~Loll, Phys. Reports 519 (2012) 127.


\bibitem{unparticles}
 H.~Georgi, Phys. Rev. Lett. 98 (2007) 221601; Phys. Lett. B650 (2007) 275-278.


\bibitem{oliver+wetterich+mr}
 O.~Lauscher, M.~Reuter and C.~Wetterich,  Phys. Rev. D 62 (2000) 125021.

\bibitem{lasagne}
A.~Bonanno and M.~Reuter, 
Phys. Rev. D 87 (2013) 8, 084019.

\bibitem{cwRGimpprop}
 C.~Wetterich, Z.~Phys. C57 (1993) 451.



\bibitem{powers}
 C.~G.~Bollini and J.~J.~Giambiagi, J. Math. Phys. 34 (1993) 610;
D.~L\'{o}pez Nacir and F.~D.~Mazzitelli, Phys.~Rev. D75 (2007) 024003.



\bibitem{tohline}
 J.~E.~Tohline, in {\it IAU Symposium 100, Internal Kinematics and Dynamics of Galaxies}, edited
by E. Athanassoula, Reidel, Dordrecht, 1983, p. 205.


\bibitem{kuhn}
 J.~R.~Kuhn and L.~Kruglyak, Astrophys. J. 313, 1 (1987).



\bibitem{bekenstein-review}
 J.~D.~Bekenstein, in {\it Second Canadian Conference on General Relativity and Relativistic Astrophysics}, A. Coley, C. Dyer and T. Tupper, Eds., World Scientific, Singapore, 1988, p. 68.



\bibitem{hehl-mashhoon}
 F.~W.~Hehl and B.~Mashhoon, 
Phys. Lett. B 673, 279 (2009); \\
Phys. Rev. D 79, 064028 (2009);\\
H.-J.~Blome, C.~Chicone, F.~W.~Hehl and B.~Mashhoon,\\ Phys. Rev. D 81, 065020 (2010).

\bibitem{tully}
 R.~B.~Tully and J.~R.~Fisher, Astron.~Astrophys. 54 (1977) 661-673.


\bibitem{mashhoon1}

C.~Chicone and B.~Mashhoon, J.~Math.~Phys. 53 (2012) 042501.


\bibitem{mashhoon2}

S.~Rahvar and B.~Mashhoon, Phys. Rev. D89 (2014) 104011.


\bibitem{hehl-tele}
 M.~Blagojevi\'{c} and F.~W.~Hehl, {\it Gauge Theories of Gravitation},\\ Imperial College Press, London, 2013.


\bibitem{dr1}
 W.~Dittrich and M.~Reuter, {\it Effective Lagrangians in Quantum Electrodynamics}, Springer, Berlin, 1985.


\bibitem{andi-MG}
A.~Nink and M.~Reuter, Int. J. Mod. Phys. D 22, 1330008 (2013), \mbox{arXiv:1212.4325}.


\bibitem{cosmo1}
A.~Bonanno and M.~Reuter,
Phys.\ Rev.\ D 65 (2002) 043508, \mbox{hep-th/0106133};\\ M.~Reuter and 
F.~Saueressig, JCAP\ 09 (2005) 012, \mbox{hep-th/0507167}.


\bibitem{weinberg-AS-infl}
 S.~Weinberg, Phys. Rev. D81 (2010) 083535.

\bibitem{alfio-As-infl}
 A.~Bonanno, Phys. Rev. D85 (2012) 081503.


\bibitem{Naxos-book}
G.~Calcagni et al. (Eds.), {\it Proceedings of the 6th Aegean Summer School on Quantum gravity and quantum cosmology}, Lect. Notes Phys. 863, Springer (2013).


\bibitem{planck}

Planck Collaboration - P.A.R.~Ade, et al., \mbox{arXiv:1303.5062} [astro-ph.CO].


\bibitem{bicep}

BICEP2 Collaboration - P.A.R.~Ade, et al., \mbox{arXiv:1403.3985} [astro-ph.CO].



\end{thebibliography}
\end{document}